\documentstyle[preprint,aps,floats]{revtex}
\begin{document}
%
%
\preprint{$
\begin{array}{l}
\mbox{Fermilab-Pub-98/164-T}\\[-3mm]
\mbox{PSI-PR-98-15}\\[-3mm]
\mbox{UB-HET-98-01}\\[-3mm]
\mbox{July~1998} \\ [1cm]
\end{array}
$}
\title{Electroweak Radiative Corrections to $W$ Boson Production in
Hadronic Collisions\\[1cm]}
\author{U.~Baur$^a$, S.~Keller$^b$ and D.~Wackeroth$^c$\\[0.5cm]}
\address{$^a$Department of Physics,
State University of New York, Buffalo, NY 14260, USA\\
$^b$Theory Group,
Fermi National Accelerator Laboratory, Batavia, IL 60510, USA\\
$^c$F1 Theory Group, Paul Scherrer Institute,
CH-5232 Villigen PSI, Switzerland \\}
\maketitle
\tightenlines
%
%
\begin{abstract}
\baselineskip15.pt  
The ${\cal O}(\alpha)$ electroweak radiative corrections to the process 
$p\,p\hskip-7pt\hbox{$^{^{(\!-\!)}}$} \to W^\pm\to\ell^\pm\nu$
($\ell=e,\,\mu$) are calculated. The ${\cal O}(\alpha)$ corrections 
can be decomposed into separately gauge invariant contributions to the
$W$ boson production and decay processes. Factorizing the collinear
singularity associated with initial state photon radiation into the
parton distribution functions, we find that initial state corrections
have a significantly smaller effect than final state radiative
corrections. We study in detail the effect of electroweak radiative 
corrections on a number of interesting observables: the $W$ transverse 
mass distribution, the $W$ to $Z$ transverse mass ratio, the charge 
asymmetry of leptons in $W\to\ell\nu$ decays, as well
as the $W$ production cross section and the $W$ to $Z$ cross section
ratio. We also investigate how experimental lepton identification
requirements change the effect of the electroweak corrections.
\end{abstract}
\newpage
%
%
\section{Introduction}
The Standard Model of electroweak interactions (SM)
so far has met all experimental challenges and is now tested at the 
$0.1\%$ level~\cite{holliksm}. However, there is little direct
experimental information on the mechanism  which generates the masses of
the weak gauge bosons. In the SM, spontaneous symmetry breaking is
responsible for mass generation. The existence of a Higgs boson
is a direct consequence of this mechanism. At present the negative
result of direct searches performed at LEP2 imposes a lower bound
of $M_H>87.6$~GeV~\cite{ewwg} on the Higgs boson mass. Indirect
information on the mass of the Higgs boson can be extracted from the 
$M_H$ dependence of radiative corrections to the $W$ boson mass, $M_W$,
and the effective weak mixing angle, $\sin^2\theta^{lept}_{eff}$.
Assuming the SM to be valid, a global $\chi^2$-fit to
all available electroweak precision data yields a (one-sided) 95\%
confidence level (CL) upper limit on $M_H$ of 408~GeV~\cite{caso}. 

Future more precise measurements of $M_W$ and the top quark
mass, $m_{top}$, will lead to more accurate information on the Higgs
boson mass~\cite{degrassi,Tev2000,BD}. Currently, the $W$ boson mass is 
known to $\pm
65$~MeV~\cite{jer} from direct measurements, whereas the uncertainty of 
the top quark mass is $\pm 5.2$~GeV~\cite{ward}. With a precision
of 30~MeV (10~MeV) for the $W$ mass, and 2~GeV for the top quark mass, 
$M_H$ can be predicted from a global analysis with an uncertainty of
about $30\%$ ($15\%$)~\cite{Tev2000,BD}. Comparison of these indirect 
constraints
on $M_H$ with the results from direct Higgs boson searches at LEP2, the 
Tevatron collider, and the Large Hadron Collider (LHC) will be an
important test of the SM. They will also provide restrictions on the
parameters of the Minimal Supersymmetric extension
of the Standard Model (MSSM)~\cite{hollikmssm}.

A significant improvement in the $W$ mass uncertainty is expected in the
near future from measurements at LEP2~\cite{LEPWmass} and the Fermilab
Tevatron $p\bar p$ collider~\cite{Tev2000}. The ultimate precision
expected for $M_W$ from the combined LEP2 experiments is approximately 
40~MeV~\cite{LEPWmass}. At the Tevatron, integrated luminosities of
order 1~fb$^{-1}$ are envisioned in the Main Injector Era (Run~II), and one
expects to measure the $W$ mass with a precision of approximately
50~MeV~\cite{Tev2000} per experiment. The prospects for a precise 
measurement of $M_W$ would further improve if a significant upgrade in 
luminosity beyond the goal of the Main Injector could be realized. 
With recent advances in
accelerator technology~\cite{GPJ}, Tevatron collider luminosities of
order $10^{33}\,{\rm cm}^{-2}\,{\rm s}^{-1}$ may become a reality,
resulting in integrated luminosities of up to
10~fb$^{-1}$ per year. With a total integrated luminosity of
30~fb$^{-1}$, one can target a precision of the $W$ mass of 15~--
20~MeV~\cite{Tev2000}. A similar or better accuracy may also be reached 
at the LHC~\cite{KW}.

In order to measure the $W$ boson mass with high
precision in a hadron collider environment, it is necessary to fully 
understand and control higher order QCD and electroweak (EW) corrections. A
complete calculation of the full ${\cal O}(\alpha)$ electroweak 
radiative corrections to $p\,p\hskip-7pt\hbox{$^{^{(\!-\!)
}}$} \rightarrow W^\pm \rightarrow \ell^\pm \nu$ ($\ell=e,\,\mu$) has not
been carried out yet. In a previous calculation, only the final state 
photonic corrections were included~\cite{BK,RGW}, using an 
approximation in which the sum of the soft and virtual parts is
indirectly estimated
from the inclusive ${\cal O}(\alpha^2)$ $W\to\ell\nu(\gamma)$
width and the hard photon bremsstrahlung contribution. The unknown part
of the ${\cal O}(\alpha)$ 
electroweak radiative corrections, combined with effects of multiple
photon emission (higher order corrections), have been estimated to
contribute a systematic uncertainty of $\delta M_W=
15-20$~MeV to the measurement of the $W$ mass~\cite{cdfwmass,D0Wmass}. 

In this paper we present a new and more accurate calculation of the ${\cal
O}(\alpha)$ EW corrections to resonant $W$ boson production in hadronic 
collisions. Our calculation is based on the full set of ${\cal
O}(\alpha^3)$ Feynman diagrams, and includes both initial and final 
state radiative corrections, as well as the contributions from their
interference. Final state
charged lepton mass effects are included in the following approximation. 
The lepton mass regularizes the collinear 
singularity associated with final state photon radiation. The associated
mass singular logarithms of the form $\ln(\hat s/m_\ell^2)$, where $\hat
s$ is the squared parton center of mass energy and $m_\ell$ is the
charged lepton mass, are included in our calculation, but the very small
terms of ${\cal O}(m_\ell^2/\hat s)$ are neglected. 

To perform our calculation, we use a Monte Carlo method for
next-to-leading-order (NLO) calculations similar to that described in 
Ref.~\cite{NLOMC}. With the Monte Carlo method, it is easy to calculate 
a variety of observables simultaneously and to simulate detector 
response. Calculating the EW radiative corrections to resonant $W$ boson
production, the problem arises how an unstable charged gauge boson can
be treated consistently in the framework of perturbation theory.
This problem has been studied in Ref.~\cite{dw} with 
particular emphasis on finding a gauge invariant decomposition of the 
EW ${\cal O}(\alpha)$ corrections into a QED-like and a modified weak part. 
Unlike the $Z$ boson case, the Feynman diagrams which involve
a virtual photon do not represent a gauge invariant subset. 
In Ref.~\cite{dw}, it was
demonstrated how gauge invariant contributions that contain the
infrared (IR) singular terms can be extracted from the virtual photonic 
corrections. These contributions can be combined with the also IR-singular 
real photon corrections in the soft photon region to form IR-finite 
gauge invariant QED-like contributions
corresponding to initial state, final state and interference
corrections. The collinear singularities associated with initial state
photon radiation can be removed by universal collinear counter terms
generated by ``renormalizing'' the parton distribution
functions~\cite{spies,RPS}, in complete analogy to gluon emission in
QCD. A similar strategy has been employed in a recent calculation of the
${\cal O}(\alpha)$ QED corrections to $Z$ boson production in hadronic
collisions~\cite{BKS}. 

The technical details of our calculation are described in Sec.~II. We
first extract the collinear behaviour of the partonic cross
section for both the initial and the final state corrections. Then, we 
define the quark
distribution functions in next-to-leading order QED within the QED $DIS$
and the QED $\overline{MS}$ factorization scheme
when a finite quark mass is used to regulate the 
collinear singularities. Finally, we provide explicit formulae for
the ${\cal O}(\alpha^3)$ differential cross section and perform various
consistency checks. 

Numerical results for $p\bar p$ collisions at $\sqrt{s}=1.8$~TeV are
presented in Sec.~III. In hadron collider experiments, $W$ bosons are
identified by their leptonic decays, $W\to\ell\nu$. Since the neutrino
escapes undetected, the $\ell\nu$ invariant mass cannot be
reconstructed, and one must resort to other kinematic variables for the
measurement of $M_W$. The observable which currently provides the best
measurement of $M_W$ is the distribution of the transverse mass, $M_T$.
The $M_T$ distribution sharply peaks at $M_W$, and is rather
insensitive to QCD corrections~\cite{mt}. 
Alternative measurements of the $W$ mass are
provided~\cite{D0Wmass} by the lepton $p_T$ distribution which peaks at
$M_W/2$, and the $W/Z$ transverse mass ratio~\cite{GK,srini}. Due to the
mass singular logarithms associated with final state photon
bremsstrahlung in the limit where the photon is emitted collinear with
the charged lepton, the distributions which are sensitive to $M_W$, the
$W$ production cross section and the $W$ to $Z$ cross section ratio in
presence of cuts, and the charge asymmetry of leptons in $W$ decays are
significantly affected by the ${\cal O}(\alpha)$ electroweak radiative
corrections. 

The size of the radiative corrections strongly depends on the 
detector resolution. In Sec.~III, using a simplified model of the D\O\ 
detector as an example, we also investigate how the finite resolution of
realistic detectors affects the electroweak 
radiative corrections. Electrons and photons which are almost
collinear are difficult to discriminate, and the momenta of the two
particles are thus recombined into an effective electron
momentum~\cite{cdfwmass,D0Wmass} if their separation in the
pseudo-rapidity -- azimuthal angle plane is below a critical value. This
procedure completely eliminates the mass singular logarithms and,
therefore, strongly reduces the size of the ${\cal O}(\alpha)$
corrections. In
contrast, photons which are almost collinear with muons are rejected if 
they are too energetic~\cite{cdfwmass,D0Wcross} which results in
residual mass singular logarithmic
corrections to observable quantities in $W$ production. Finally, our
conclusions are presented in Sec.~IV.

\section{${\cal O}(\alpha)$ Electroweak Corrections to $W$ Production in
Hadronic Collisions}

\begin{figure}
\vskip 18cm
\includegraphics{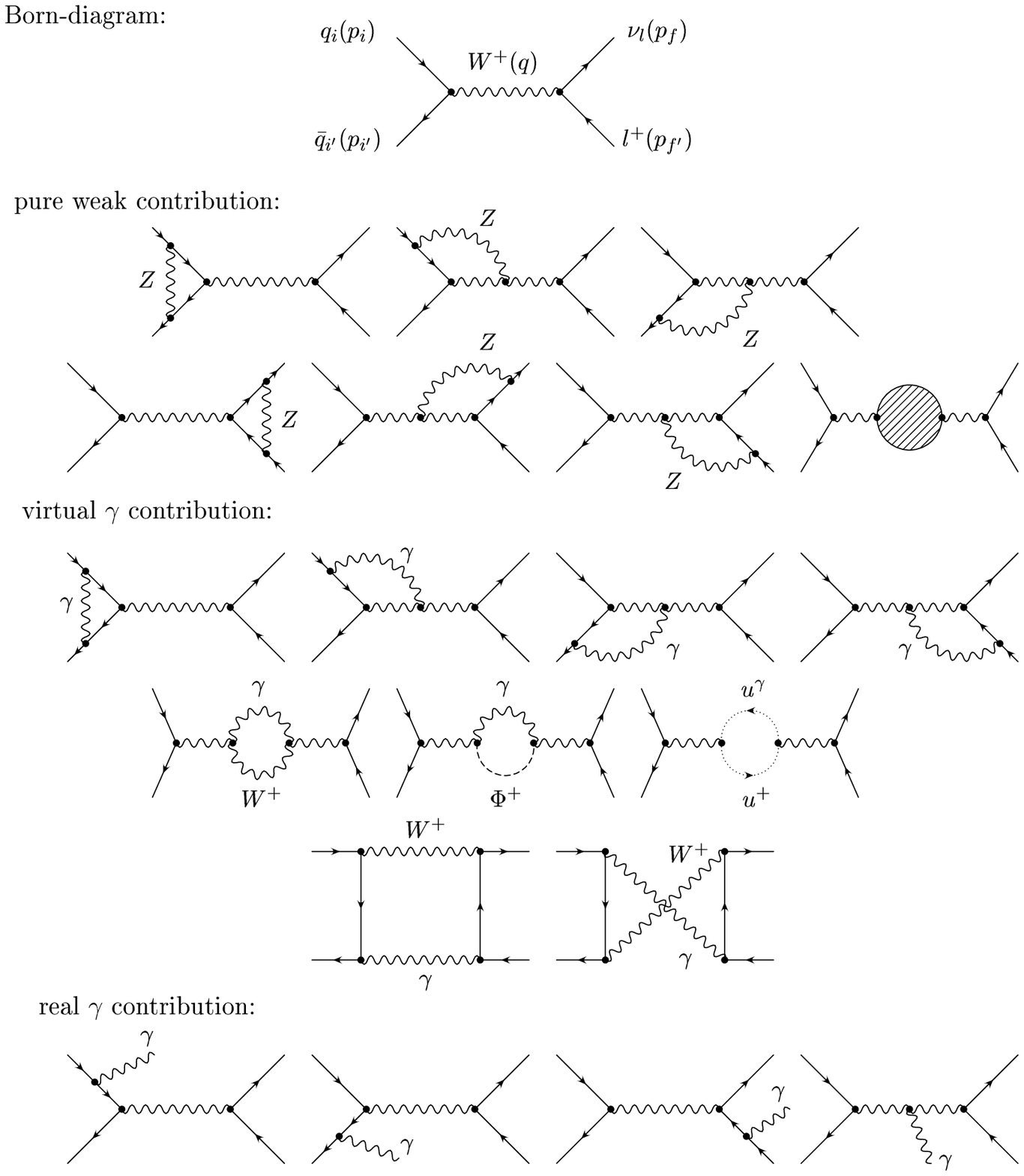}
\caption{The Feynman diagrams contributing to $W$ boson production at
${\cal O}(\alpha^3)$ ($\Phi^+$: Higgs~-- ghost field, 
$u^+,u^{\gamma}$: Faddeev-Popov-ghost fields; the non-photonic contribution
to the $W$ self energy insertion is symbolized by the shaded loop).  
An explicit representation of the non-photonic contribution
to the $W$ self energy insertion can be found in Ref.~[\ref{fort}].}
\label{fig:one}
\end{figure}

The calculation presented here employs a
combination of analytic and Monte Carlo integration
techniques\footnote{A parton level Monte Carlo program (in FORTRAN) for 
$p\,p\hskip-7pt\hbox{$^{^{(\!-\!)}}$} \to W^\pm\to\ell^\pm\nu$ including
${\cal O}(\alpha)$ EW corrections is available from {\tt
http://ubhex.physics.buffalo.edu/\~~$\!\!\!$baur/wgrad/wgrad.tar.gz}, or by 
contacting {\tt baur@ubhex.physics.buffalo.edu}.}. Details
of the method can be found in Ref.~\cite{NLOMC}.
The Feynman diagrams contributing to $W$ boson production in hadronic
collisions to ${\cal O}(\alpha^3)$,
\[ 
q_i (p_i) \overline{q}_{i'} (p_{i'}) \rightarrow  W^+ (q)(\gamma)
\rightarrow \nu_\ell (p_f) \ell^+ (p_{f'}) (\gamma(k))
\]
are shown in Fig.~\ref{fig:one}. Since we are interested in the cross 
sections in the vicinity of the $W$ resonance, the $W,Z$ box diagrams
can be neglected as non-resonant contributions of higher order in 
perturbation theory, and thus are not depicted in Fig.~\ref{fig:one}.
The calculation of
$\ell\nu$ production in hadronic collisions at ${\cal O}(\alpha^3)$
includes contributions from the square of the Born graphs, the
interference between the Born diagrams and the virtual one loop graphs,
and the square of the real emission diagrams. 

Our treatment of the ${\cal O}(\alpha)$ corrections to $W$
boson production in the resonance region is based on the calculation 
presented in Ref.~\cite{dw}, which we outline here. Unlike the
$Z$ boson case, the
Feynman diagrams of Fig.~\ref{fig:one} which involve a virtual photon do
not represent a gauge invariant subset. In Ref.~\cite{dw}, it was
demonstrated how gauge invariant contributions that contain the
infrared singular terms can be extracted from the virtual photonic 
corrections. These contributions can be combined with the also IR-singular 
real photon corrections in the soft photon region to form IR-finite 
gauge invariant QED-like contributions
corresponding to initial state, final state and interference
corrections. The soft photon region is defined by requiring that the
photon energy in the parton center of mass frame, $\hat E_{\gamma}$, is
$\hat E_\gamma<E_{cut}=\delta_s\sqrt{\hat s}/2$.  In this phase
space region, the soft photon approximation can be used to calculate
the cross section, provided that $\delta_s$ is sufficiently small.
The soft singularities are regularized by giving the photon a fictitious
small mass.  In the sum of the
virtual and soft photon terms the unphysical photon mass dependence
cancels, and the QED-like contributions are IR finite.
 
The IR finite remainder of the virtual photonic corrections and the
pure weak one-loop corrections of Fig.~\ref{fig:one} 
can be combined to separately
gauge invariant modified weak contributions to the $W$ boson
production and decay processes.  Both the QED-like and the modified
weak contributions are expressed in terms of form factors,
$F_{Q\!E\!D}^{a}$ and $\tilde F_{weak}^{a}$, which multiply the Born
cross section~\cite{dw}. The superscript $a$ in the form factors
denotes the initial state, final state or interference contributions.

The complete ${\cal O}(\alpha^3)$ parton level cross section of
resonant $W$ production via the Drell-Yan mechanism $q_i
\overline{q}_{i'}\to\ell^+\nu(\gamma)$ can then be written as
follows~\cite{dw}:
\begin{eqnarray}\label{eq:one}
d \hat{\sigma}^{(0+1)} & = &
d \hat \sigma^{(0)}\; [1+ 2 {\cal R}e (\tilde F_{weak}^{initial}+
\tilde F_{weak}^{final})(M_W^2)] 
\nonumber \\[1.mm]
&+ & \sum_{a=initial,final,\atop interf.} [d\hat\sigma^{(0)}\; 
F_{Q\!E\!D}^a(\hat s,\hat t)+
d \hat \sigma_{2\rightarrow 3}^a] \; ,
\end{eqnarray}
where the Born cross section, $d \hat \sigma^{(0)}$, is of
Breit-Wigner form and $\hat s$ and $\hat t$ are the usual Mandelstam
variables in the parton center of mass frame.  The modified weak
contributions have to be evaluated at $\hat
s=M_W^2$~\cite{dw}. Explicit expressions for the form factors
$F_{Q\!E\!D}^a,\tilde F_{weak}^a$ are given in Ref.~\cite{dw}.  
The IR finite contribution $d\hat\sigma_{2\rightarrow 3}^a$ describes real
photon radiation with $\hat E_{\gamma}>E_{cut}$.

Additional singularities occur when the photon is collinear
with one of the charged fermions.  These collinear singularities are
regularized by retaining finite fermion masses.  Thus, both $d
\hat \sigma^{a}_{2\rightarrow 3}$ and $F_{Q\!E\!D}^{a}$ ($a=initial,
final$) contain large mass singular logarithms which have to be
treated with special care.  In the case of final state photon
radiation, the mass singular logarithms cancel when inclusive
observables are considered (KLN theorem~\cite{kln}).  For exclusive
quantities, however, these logarithms can result in large corrections,
and it may be necessary to perform a resummation of the soft and/or
collinear photon emission terms. To increase the numerical stability of 
the inclusive calculation, it is advantageous to extract the collinear 
part from $d\hat \sigma^{final}_{2\rightarrow 3}$ and perform the 
cancellation of the mass singular logarithms analytically. The reduced
$2 \rightarrow 3$ contribution, {\it i.e.} the real photon 
contribution away from the soft and collinear region, can be evaluated 
numerically using standard Monte Carlo techniques. 

For initial state photonic
corrections, the mass singular logarithms always survive. These
logarithmic terms are equivalent to the $1/\epsilon$ singularity
encountered in dimensional regularization ($D=4-2\epsilon$ is the number
of dimensions) with massless quarks. They are
universal to all orders in perturbation theory, and can therefore be
canceled by universal collinear counter terms generated by
`renormalizing' the parton distribution functions (PDF's), in complete
analogy to gluon emission in QCD\footnote{Alternatively, these
logarithmic terms can be retained in the calculation. They would lead to
large corrections, but then also to large changes in the input PDF's.}. 
In addition to the collinear counterterms, finite terms can be absorbed
into the PDF's, introducing a QED factorization scheme dependence. We
have carried out our calculation in the QED $DIS$ and QED $\overline{MS}$
scheme. The extraction of the collinear part
from $d \hat \sigma^a_{2\rightarrow 3}$ and the renormalization
of the PDF's are described in Sec.~IIA and~IIB, respectively. 
In Sec.~IIC we provide explicit expressions for the ${\cal
O}(\alpha^3)$ cross section for $W$ production in hadronic collisions in
the QED $DIS$ and QED $\overline{MS}$ scheme,
and study the dependence of the ${\cal O}(\alpha^3)$ cross section on
the theoretical cutoff parameters which define the soft and collinear
regions. 

\subsection{The Extraction of Collinear Singularities from 
$d \hat \sigma^{a}_{2\rightarrow 3}$}

The contribution of real photon emission to the ${\cal O}(\alpha^3)$ 
cross section for $W$ production in hadronic collisions is given by
\begin{equation}\label{eq:two}
d\hat \sigma_{real}=d P_{2\rightarrow 3}~ \overline{\sum}
|{\cal M}_{B\!R}|^2 ,
\end{equation}
where $d P_{2\rightarrow 3}$ is the product of the three particle phase 
space element and the flux factor,
\begin{equation}\label{eq:three}
d P_{2\rightarrow 3}=\frac{1}{2\hat s}\, \frac{1}{(2\pi)^5}\,
\frac{d^3 p_f d^3 p_{f'} d^3 k}{8 p_f^0 p_{f'}^0 k^0}\, \delta(p_i+p_{i'}-p_f
-p_{f'}-k),
\end{equation}
and the Bremsstrahlung matrix element ${\cal M}_{B\!R}$ is given by
\begin{eqnarray}\label{eq:four}
{\cal M}_{B\!R}&  = & i\, \frac{\pi \alpha}{2 s_w^2}
 \; \sqrt{4\pi\alpha}
\;  \left\{\frac{1}{\hat s-M_W^2} \;
\overline{u}_f\,  G_{\mu,f}^{\rho}\, (1-\gamma_5) \,
v_{f'}\, \overline{v}_{i'} \gamma^{\mu}(1-\gamma_5)u_i
\right. \nonumber\\
& -& \left. \frac{1}{\hat s-M_W^2-2kq}\; \overline{u}_f \, 
\gamma_{\mu}(1-\gamma_5)\, v_{f'} \, \overline{v}_{i'} \,
G_i^{\mu\rho}(1-\gamma_5)\, u_i \right\} \; \epsilon^{\ast}_{\rho}(k) \; .
\end{eqnarray}
In Eq.~(\ref{eq:four}), $s_w^2=\sin^2\theta_W$, where $\theta_W$ is the
weak mixing angle, $\epsilon_{\rho}$ denotes the photon polarization 
vector, and
\begin{eqnarray}\label{eq:five}
G^{\mu\rho}_f & = & 
Q_f\, \frac{(p_f^{\rho}+\gamma^{\rho}  \not\!{k}/2)\, \gamma^{\mu}}
{kp_f}-Q_{f'}\, \frac{\gamma^{\mu}\, (p_{f'}^{\rho}+\not\!{k}\, 
\gamma^{\rho}/2)}
{kp_{f'}}-\frac{\gamma^{\mu}\, q^{\rho}+k^{\mu}\, \gamma^{\rho}
-g^{\mu\rho}\, \not\!{k}}{kq}\, ,
\nonumber\\[2.mm]
G^{\mu\rho}_i & = & Q_i\, \frac{\gamma^{\mu}\, (p_i^{\rho}
-\not\!{k}\, \gamma^{\rho}/2)}
{kp_i}-Q_{i'}\, \frac{(p_{i'}^{\rho}-\gamma^{\rho} \not\!{k}/2)\, 
\gamma^{\mu}}
{kp_{i'}}-\frac{\gamma^{\mu}\, q^{\rho}-k^{\mu}\,  \gamma^{\rho}
+g^{\mu\rho}\, \not\!{k}}{kq} \, .
\end{eqnarray}
$Q_a$ ($a=i,\,i',f,\,f'$) denotes the electric charge in units of the
proton charge, $e$. 
The initial and final state currents are separately conserved:
$k_{\rho} \, G^{\mu\rho}_f=(Q_f-Q_{f'}-1)\,\gamma^{\mu}=0$ and
$k_{\rho}\, G^{\mu\rho}_i=(Q_i-Q_{i'}-1)\,\gamma^{\mu}=0$. 

$d\hat \sigma_{real}$ can be decomposed into soft and hard initial
state, final state and interference terms:
\begin{equation}
d\hat \sigma_{real}= \sum_{a=initial,final,\atop interf.}
(d\hat \sigma_{soft}^a+ d\hat\sigma^a_{2\rightarrow 3}).
\end{equation}
Here, $d\hat \sigma_{soft}^a$ are the soft photon contributions  
($\hat E_\gamma<E_{cut}$) and, as explained before, are included in the 
QED-like form factors $F_{Q\!E\!D}^a$. 

The initial and final state hard
photon contributions, $d\hat \sigma^{initial}_{2\rightarrow 3}$ and 
$d\hat \sigma^{final}_{2\rightarrow 3}$, contain mass singular 
logarithmic terms, whereas the interference contribution, $d\hat 
\sigma^{interf.}_{2\rightarrow 3}$, does not.
In order to extract the mass singular terms from $d\hat 
\sigma^{initial}_{2\rightarrow 3}$ and $d\hat 
\sigma^{final}_{2\rightarrow 3}$, we define a collinear region by 
requiring that 
\begin{equation}
\cos\theta>1-\delta_{\theta},
\end{equation}
where $\theta$ is the angle between the charged fermion and the emitted
photon in the parton center of mass frame. $d\hat \sigma_{2\rightarrow 
3}^a$ can be decomposed into a finite contribution away 
from the soft and collinear singularity, $d\hat 
\sigma^{a,finite}_{2\rightarrow 3}$, which will be evaluated numerically,
and a collinear part $d\hat\sigma_{coll.}^a$, for which the integration over
the singular phase space region can be performed analytically,
\begin{equation}\label{eq:six}
d \hat\sigma_{2\rightarrow 3}^a = 
d \hat\sigma_{coll.}^a +d \hat\sigma_{2\rightarrow 3}^{a,finite} 
\qquad a=initial,~final .
\end{equation}
In the following we calculate $d\hat \sigma_{coll.}^a$ explicitly
for both initial and final state photon radiation.

\begin{figure}
\vskip 7cm
\includegraphics{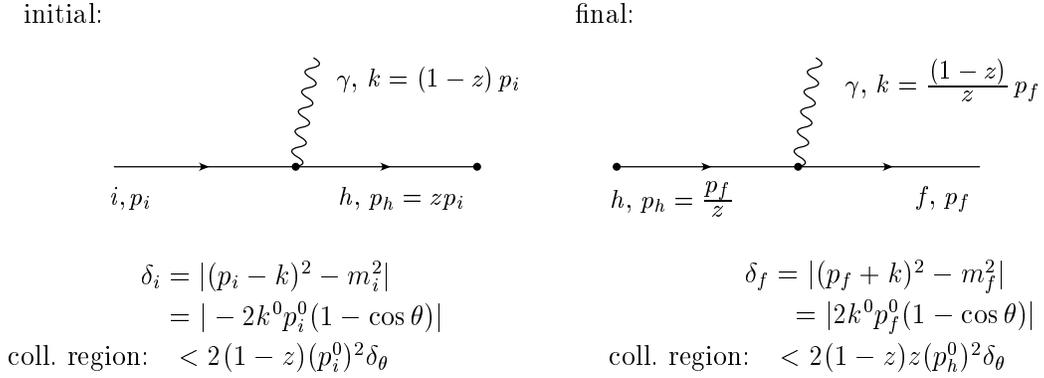}
\vskip -1.cm
\caption{The splitting of the initial and final state fermions 
$i(p_i)\rightarrow h(p_h)+\gamma$ and $h(p_h)\rightarrow f(p_f)+\gamma$ 
in the collinear region. The 
hard momentum $p_h$ represents the amount of the parent momentum $p_{i,f}$
after (before) the emission of a collinear photon.}
\label{fig:two}
\end{figure}

In the collinear region, $dP_{2\rightarrow 3}$ factorizes into a two particle
and a collinear part (see Fig.~\ref{fig:two} for notation)\\[4.mm]
\vbox{
\noindent\underline{initial state:}
\vskip -2.mm
\begin{equation}\label{eq:seven}
dP_{2\rightarrow 3}(i+i'\rightarrow f+f'+\gamma)\rightarrow  
dP_{2\rightarrow 2}(h+i'\rightarrow f + f')\; 
\frac{z d^3 k}{2 (2\pi)^3 k^0} = -dP_{2 \rightarrow 2}\;
\frac{z dz d\delta_i}{16\pi^2}
\end{equation}
\vskip -3.mm
}
\vbox{
\noindent\underline{final state:}
\vskip -2.mm
\begin{equation}\label{eq:eight}
dP_{2\rightarrow 3}(i+i'\rightarrow f+f'+\gamma)\rightarrow  
dP_{2\rightarrow 2}(i+i'\rightarrow h + f')\;
\frac{z^2 d^3 k}{2 (2\pi)^3 k^0} = -dP_{2 \rightarrow 2}\;
\frac{dz d\delta_f}{16\pi^2} \; ,
\end{equation}
\vskip -3.mm
}
where we have used
\begin{eqnarray}\label{eq:nine} 
d^3 k & = & 2 \pi (k^0)^2 dk^0 d\cos\theta,
\nonumber\\
d\delta_{i,f} = 2k^0 p^0_{i,f}\, d\cos\theta & ; &
d k^0=-p_i^0 dz=-\frac{1}{z^2}\, p_f^0 dz  \; .
\end{eqnarray}
Using the leading pole approximation, the squared matrix element 
for initial and final state photon emission (see Eq.~(\ref{eq:four}))
factorizes into the leading-order squared matrix element, $|{\cal 
M}^{(0)}|^2$, and a collinear factor, $c_{i\rightarrow h \gamma}$ or
$c_{f\gamma\to h}$, provided the parameter $\delta_{\theta}$ is sufficiently
small: 
\begin{eqnarray}\label{eq:ten}
\overline{\sum}|{\cal M}_{B\!R}^{initial}|^2
(i+i'\rightarrow f+f'+\gamma)  & \rightarrow 
& \overline{\sum}|{\cal M}^{(0)}|^2(h+i'\rightarrow 
f+f') \; c_{i\rightarrow h \gamma}~, \nonumber \\[2.mm]
\overline{\sum}|{\cal M}_{B\!R}^{final}|^2
(i+i'\rightarrow f+f'+\gamma)  & \rightarrow 
& \overline{\sum}|{\cal M}^{(0)}|^2(i+i'\rightarrow 
h+f') \; c_{f\gamma\rightarrow h}~.
\end{eqnarray}
Here
\begin{eqnarray}\label{eq:eleven}
c_{i\rightarrow h \gamma} &=&
8\pi^2\; \frac{\alpha}{\pi}~Q_i^2 \; \frac{1}{\delta_i}
\; \left[ \frac{1+z^2}{z(1-z)} - \frac{2 m_i^2}{\delta_i}\right]~,
\nonumber\\[2.mm]
c_{f\gamma\rightarrow h} &=&
8\pi^2\; \frac{\alpha}{\pi}~Q_f^2 \; \frac{1}{\delta_f}
\; \left[ \frac{1+z^2}{1-z} - \frac{2 m_f^2}{\delta_f}\right]~,
\end{eqnarray}
and $m_i$ ($m_f$) is the mass of the initial (final) state fermion which
emits the photon.
Combining Eqs.~(\ref{eq:seven}), (\ref{eq:eight}) and~(\ref{eq:ten}),
the hard photon contribution in the collinear limit reads (see 
also Ref.~\cite{steph})\\[3.mm]
\underline{initial state:}
\begin{eqnarray}\label{eq:twelve}
d\hat\sigma_{coll.}^{initial} & = & 
\int_0^{1-\delta_s} dz \left\{ d\hat\sigma^{(0)}(h+i'
\rightarrow f+f') \int_{(1-z)m_i^2}^{2(1-z)(p_i^0)^2\delta_{\theta}}
d\delta_i\; \frac{z}{16\pi^2} \; c_{i\rightarrow h \gamma}
+(i\leftrightarrow i')\right\}
\nonumber\\[2.mm]
& = &  
\int_0^{1-\delta_s} dz \left\{ d\hat\sigma^{(0)} \frac{\alpha}{2\pi}\;
Q_i^2 \left[ \frac{1+z^2}{1-z}\; \ln\left(\frac{\hat s_{i'h}}{m_i^2}\,
\frac{\delta_{\theta}}{2}\,\frac{1}{z}\right)-
\frac{2z}{1-z}\right]+(i\leftrightarrow i')\right\}~,
\end{eqnarray}
\vbox{
\noindent\underline{final state:}
\begin{eqnarray}\label{eq:thirteen}
d\hat\sigma_{coll.}^{final} & = & 
\int_0^{1-\delta_s} dz \left\{ d\hat\sigma^{(0)}(i+i'
\rightarrow h+f') \int_{(1-z)/z m_f^2}^{2z (1-z)(p_h^0)^2\delta_{\theta}}
d\delta_f\; \frac{1}{16\pi^2}\; c_{f\gamma\rightarrow h}+(f\leftrightarrow
f')\right\}
\nonumber\\[2.mm]
& = &  
\int_0^{1-\delta_s} dz \left\{ d\hat\sigma^{(0)} \frac{\alpha}{2\pi}\;
Q_f^2 \left[ \frac{1+z^2}{1-z}\; \ln\left(\frac{\hat s_{f'h}}{m_f^2}\,
\frac{\delta_{\theta}}{2}\, z^2 \right)-
\frac{2z}{1-z}\right]+(f\leftrightarrow f')\right\}~,
\end{eqnarray}
\vskip -3.mm
}
with $\hat s_{i'h;f'h}= (p_h + p_{i';f'})^2$. In order to avoid double
counting in the soft region, the upper limit in the $z$ integration has
to be reduced from $z=1$ to $z=1-\delta_s$ in Eqs.~(\ref{eq:twelve})
and~(\ref{eq:thirteen}).  

\subsection{Mass Factorization: QED $DIS$ and $\overline{MS}$ Scheme}

The mass singular logarithmic terms of Eq.~(\ref{eq:twelve}) 
can be absorbed by the counter terms to the PDF's. In addition to the 
singular terms, finite ${\cal O}(\alpha)$ terms can be absorbed into the
PDF's. At next-to-leading order (NLO) in QED, the parton distribution 
functions therefore depend on the QED factorization scheme used. In this
subsection, we derive the NLO PDF's in the QED $DIS$ and the QED 
$\overline{MS}$ scheme.

\begin{figure}
\vskip 7.cm
\includegraphics{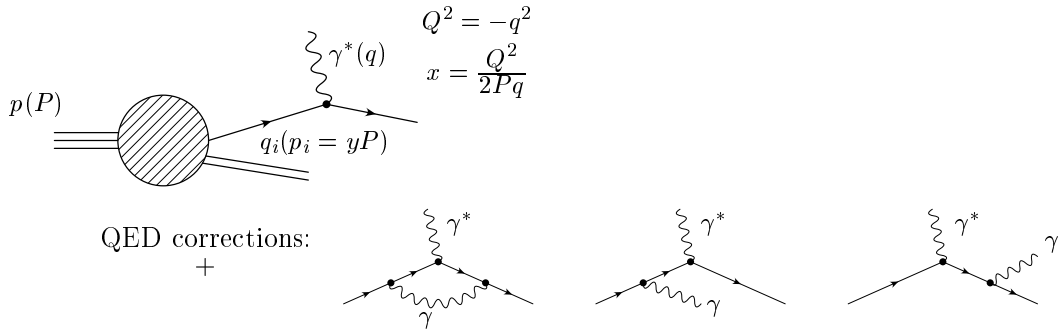}
\vskip -3.5cm
\caption{The QED one-loop corrections in deep inelastic lepton -- 
nucleon scattering.}
\label{fig:three}
\end{figure}

In order to derive
the parton distribution functions at next-to-leading order in QED, one
must calculate the virtual and real photon contribution to the square of
the parton electromagnetic current, integrated over the phase space of
the final state partons, $\widehat W^i_{\mu\nu}$. The contributing
Feynman diagrams are shown
in Fig.~\ref{fig:three}. The tensor $\widehat W^i_{\mu\nu}$ is related
to the structure function $F_2(x,Q^2)$ by (see Fig.~\ref{fig:three} for
notation)
\begin{equation}\label{eq:fifteen}
F_2(x,Q^2) = \frac{x}{4\pi}\sum_i \int_x^1 \frac{dy}{y} \; q_i(y)
\left[ -g^{\mu\nu}+\frac{12x^2}{Q^2 y^2} \; P^{\mu} P^{\nu}\right]\; 
\widehat W^i_{\mu\nu} \; ,
\end{equation}
where the sum is taken over all contributing quark flavors, and the
$q_i(y)$ are the unrenormalized quark distribution functions. Since it 
involves an additional power of $\alpha$, 
we do not take into account the photon content of the proton in our
calculation. 

In the physical ($DIS$) scheme~\cite{OWENSTUNG}  the `renormalized' quark 
distribution functions are defined by requiring that $F_2(x,Q^2)$ is 
given by the sum of the quark distributions to all orders in 
perturbation theory
\begin{equation}
F_2(x,Q^2)=x \sum_i Q_i^2 \left [q_i^{DIS}(x,Q^2)+\bar q_i^{DIS}(x,Q^2)
\right]~,
\label{eq:f2dis}
\end{equation}
where the QED factorization scale has been set equal to $Q$.
The ${\cal O}(\alpha)$ structure function $F_2(x,Q^2)$ can be obtained
from the corresponding ${\cal O}(\alpha_s)$ QCD structure 
function~\cite{paige} by the replacement
\begin{equation}\label{eq:fourteen}
\frac{\alpha_s}{\pi}\,  \frac{4}{3} \rightarrow \frac{\alpha}{\pi}\,
Q_i^2 \; .
\end{equation}
For massive fermions one finds in the limit $Q^2\gg m_i^2$ ($z=x/y$):
\begin{eqnarray}\label{eq:sixteen}
F_2(x,Q^2) & = & x\sum_i Q_i^2 \left\{ \int_x^1 \frac{dz}{z}
\; q_i\left(\frac{x}{z}\right)\left\{\delta(1-z)+\frac{\alpha}{\pi}\, Q_i^2
\left[ \left(2\ln\delta_s+\frac{3}{2}\right)\ln\left(\frac{Q^2}{m_i^2}\right)
\delta(1-z)\right. \right. \right.
\nonumber\\[2.mm]
&+& \left. \left. \left. 
\frac{1+z^2}{1-z}\,\ln\left(\frac{Q^2}{m_i^2}\,\frac{1}{(1-z) z}
\right)\theta(1-\delta_s-z) \right. \right. \right. \nonumber\\[2.mm]
&-& \left. \left. \left.\left[\ln^2 \delta_s
+\frac{7}{2}\, \ln \delta_s+\frac{5}{2} 
+ \frac{\pi^2}{3} \right] \delta(1-z)
+\left[\frac{1}{2}\, \frac{1-8 z}{1-z}
+3 z\right] \theta(1-\delta_s-z)\right]\right\}\right\}~.
\end{eqnarray}
Using Eq.~(\ref{eq:sixteen}) it is then straightforward to calculate 
$q_i^{DIS}(x,Q^2)$ in terms of the unrenormalized quark distribution 
functions $q_i(x)$ (see below).
The relation between $F_2(x,Q^2)$ and the quark distribution functions
in the $\overline{MS}$ scheme~\cite{MSBAR} is given by
\begin{eqnarray}
F_2(x,Q^2)&=& x \sum_i Q_i^2 \left [q_i^{\overline{MS}}(x,Q^2)
+\bar q_i^{\overline{MS}}(x,Q^2)\right ]
\nonumber\\[2.mm]
&+& x \sum_i Q_i^2 \int_x^1 \frac{dy}{y}\, \left [q_i^{\overline{MS}}(y,Q^2)
+\bar q_i^{\overline{MS}}(y,Q^2)\right ]\, \frac{\alpha}{\pi}\,
c_i\!\!\left(\frac{x}{y}\right)~,
\label{eq:f2ms}
\end{eqnarray} 
with~\cite{will}
\begin{eqnarray}
c_i(z) & = & \frac{1}{2}\, Q_i^2 \left\{
\left[\ln^2 \delta_s-\frac{3}{2}\, \ln \delta_s
- \frac{9}{2} + \frac{\pi^2}{3} \right] \delta(1-z) \right.
\nonumber\\[2.mm]
&+& \left. \left[\frac{1+z^2}{1-z}\, \ln\frac{1-z}{z}
-\frac{3}{2} \frac{1}{1-z}+2 z+3\right] \theta(1-\delta_s-z)\right\}~.
\end{eqnarray}
$c_i(z)$ represents the finite part of the QED $O(\alpha)$ corrections 
to deep inelastic scattering after removing the singularities according 
to the $\overline{MS}$ prescription. The $\overline{MS}$ scheme is defined
in the framework of dimensional regularization but Eq.~(\ref{eq:f2ms})
can also be used for its definition. To obtain the
`renormalized' quark distribution functions, $q_i^{\overline{MS}}(x,Q^2)$, 
when finite quark masses are used as regulators, we make use of the 
relation
\begin{equation}
q_i^{\overline{MS}}(x,Q^2) +\int_x^1 \frac{dy}{y}\,
q_i^{\overline{MS}}(y,Q^2)\, \frac{\alpha}{\pi}\, 
c_i\!\!\left(\frac{x}{y}\right)=q_i^{DIS}(x,Q^2)~,
\end{equation}
which follows from Eqs.~(\ref{eq:f2dis}) and~(\ref{eq:f2ms}).

The final expression for the scheme dependent `renormalized' quark 
distribution function in NLO QED is 
\begin{eqnarray}\label{eq:nineteen}
q_i(x,Q^2) &=& q_i(x) \; \left[1+\frac{\alpha}{\pi} \; Q_i^2 
\left\{1-\ln\delta_s -\ln^2\delta_s + \left(\ln\delta_s
+\frac{3}{4}\right)\,\ln\left(\frac{Q^2}{m_i^2}\right)
-\frac{1}{4} \lambda_{F\!C} f_{v+s}\right\} \right]
\nonumber\\[2.mm]
&+& \int_x^{1-\delta_s} \frac{d z}{z}\; q_i\left(\frac{x}{z}\right)
\; \frac{\alpha}{2 \pi} \; Q_i^2
\left\{\frac{1+z^2}{1-z} 
\ln\left(\frac{Q^2}{m_i^2}\frac{1}{(1-z)^2}\right)
-\frac{1+z^2}{1-z}+\lambda_{F\!C} f_c \right\}~,
\end{eqnarray}
with
\begin{equation}\label{eq:twenty}
f_{v+s} = 9+\frac{2 \pi^2}{3}+3\ln\delta_s-2 \ln^2\delta_s~ ,
\end{equation}
and
\begin{equation}\label{eq:twentyone}
f_c = \frac{1+z^2}{1-z} \ln\left(\frac{1-z}{z}\right)
-\frac{3}{2}\frac{1}{1-z}+2 z+3 \; .
\end{equation}
The QED $DIS$ ($\overline{MS}$) scheme corresponds to $\lambda_{F\!C}=1$
($\lambda_{F\!C}=0$). 

\subsection{The Cross Section for $p\,p\hskip-7pt\hbox{$^{^{(\!-\!)}}$} 
\rightarrow W(\gamma)\rightarrow\ell\nu(\gamma)$}

The differential cross section for 
$p\,p\hskip-7pt\hbox{$^{^{(\!-\!)}}$} \rightarrow W(\gamma) \rightarrow 
\ell\nu(\gamma)$ is obtained by convoluting the parton cross
section of Eq.~(\ref{eq:one}) with $q_i(x)$, and subsequently replacing 
the unrenormalized quark distribution functions by $q_i(x,Q^2)$,
using Eq.~(\ref{eq:nineteen}). The initial state QED-like contribution 
$d \hat\sigma^{(0)} F_{Q\!E\!D}^{initial}$
and the collinear part $d\hat \sigma_{coll.}^{initial}$,
including the effect of mass factorization, can be grouped into a single 
$2 \rightarrow 2$ contribution:
\begin{eqnarray}\label{eq:twentytwo}
d\sigma^{initial}_{2\rightarrow 2} & = & \sum_{i,i'}
\int dx_1 dx_2 \; [q_i(x_1,Q^2) \; \overline{q}_{i'}(x_2,Q^2)
\; d\hat \sigma^{(0)} +(1 \leftrightarrow 2) ]
\nonumber\\[2.mm]
&\times & \frac{\alpha}{\pi}
\left\{(Q_i^2+Q_{i'}^2)\left[\left(\ln\delta_s+\frac{3}{4}\right)
\ln\left(\frac{\hat s}{Q^2}\right)+\frac{\pi^2}{6} -2 +\ln^2\delta_s 
+ \frac{1}{4}\lambda_{F\!C} \;f_{v+s}\right] \right.
\nonumber\\[2.mm]
&-& \left. \ln\delta_s+\frac{3}{2}+\frac{\pi^2}{24}\right\}
\nonumber\\[2.mm]
&+ & \sum_{i,i'} \int dx_1 dx_2 \left\{\int_{x_2}^{1-\delta_s} 
\frac{d z}{z}\; 
\left[ Q_i^2  \, q_i\!\left(\frac{x_2}{z},Q^2\right) \,
\overline{q}_{i'}(x_1,Q^2)+Q_{i'}^2\,
q_i(x_1,Q^2) \,\overline{q}_{i'}\!\left(\frac{x_2}{z},Q^2\right)\right]
\right.
\nonumber\\[2.mm]
& \times & \left. d\hat\sigma^{(0)} \; \frac{\alpha}{2\pi} \;
\left\{\frac{1+z^2}{1-z}\,\ln\left(\frac{\hat s}{Q^2}\,\frac{(1-z)^2}{z}\, 
\frac{\delta_{\theta}}{2}\right)
+1-z-\lambda_{F\!C} f_c \right\} +(1 \leftrightarrow 2) \right\}  \; .
\end{eqnarray}
As expected, the mass singular logarithms cancel completely. $x_1$ and
$x_2$ in Eq.~(\ref{eq:twentytwo}) are the momentum fractions of the
parent hadrons carried by the partons.

In order to treat the ${\cal O}(\alpha)$ initial state QED-like
corrections to $W$ production in hadronic collisions in a
consistent way, QED corrections should be incorporated in the global 
fitting of the PDF's using the same factorization scheme which has been
employed to calculate the cross section. Current 
fits~\cite{PDF1} to the PDF's do not include QED corrections. A
study of the effect of QED corrections on the evolution of the parton 
distribution functions indicates~\cite{spies} that the modification 
of the PDF's is small. We have not attempted to include QED
corrections to the PDF evolution in the calculation presented here. The 
missing QED corrections to the PDF introduce an uncertainty which,
however, is likely to be smaller than the present uncertainties on 
the parton distribution functions. 

The squared matrix elements for different
QED factorization schemes differ by the finite ${\cal O}(\alpha)$ terms
which are absorbed into the PDF's in addition to the singular terms. 
In the QED $DIS$ scheme, the contribution of the QED-like initial state 
corrections to the cross section is about $8\%$ 
smaller than in the QED $\overline{MS}$ scheme. The factorization 
scheme dependence is expected to be reduced when the ${\cal O}(\alpha)$ 
QED corrections to the PDF are included. In the following, for the numerical 
evaluation of $d\sigma^{initial}_{2\rightarrow 2}$, we use the QED 
$\overline{MS}$ scheme. 

The final state $2\rightarrow 2$ contribution can be obtained directly
from the form factor $F_{Q\!E\!D}^{final}$ of Ref.~\cite{dw} 
with $Q_{f=\nu}=0$ and $Q_{f'=\ell}=-1$:
\begin{eqnarray}\label{eq:twentythree}
d\sigma_{2 \rightarrow 2}^{final} & = & \sum_{i,i'}
\int dx_1 dx_2 \; [q_i(x_1,Q^2) \; \overline{q}_{i'}(x_2,Q^2)
 \; d\hat \sigma^{(0)} +(1 \leftrightarrow 2) ]
\nonumber\\
&\times & \frac{\alpha}{\pi} \left\{\left(\ln \delta_s+\frac{3}{4}\right) 
\ln\left(\frac{\hat s}{m_\ell^2}\right)-2\ln\delta_s+\frac{1}{2}+\frac{5 
\pi^2}{24}\right\}\;  .
\end{eqnarray}
In sufficiently inclusive observables the mass singular logarithmic
terms cancel in the sum of $d\sigma_{2\rightarrow 2}^{final}$
and $d\sigma_{2 \rightarrow 3}^{final}$. 

The approximation used so far in modeling the EW radiative corrections to
$W$ boson production at the Tevatron~\cite{BK,RGW} ignores all weak,
interference and 
initial state photonic corrections, and differs from our calculation in
the treatment of the final state virtual and soft photon
contribution. At the parton level, the difference between
Eq.~(\ref{eq:twentythree}) and the $2\to 2$ contribution to the
differential cross section in the approximate calculation, is given by
\begin{eqnarray}\label{eq:twentyfive}
\Delta \hat \sigma^{final} & = & d\hat \sigma^{(0)}\;
\frac{\alpha}{2 \pi} \left\{
\left(\ln\left(\frac{m_\ell^2}{M_W^2}\, \delta_s^2\right)+\frac{7}{2}\right)
\; \ln\left(\frac{\hat s}{M_W^2}\right)+\frac{3 \pi^2}{4}-1 \right\} \; .
\end{eqnarray}
In Sec.~IIIA we shall demonstrate that the difference has a 
non-negligible effect on the shape of the transverse mass distribution. 

Experimentally, photons which are collinear with
muons can be identified without problems: photons deposit energy in the
electromagnetic calorimeter, whereas muons are identified by hits in the
muon chambers. For muons in the final state, therefore, one has to 
retain full information on the particle momentum four vectors. 
In the electron case, on the other hand, the finite resolution of 
detectors makes it difficult to discriminate between electrons and 
photons with a small opening angle, and the electron and photon 
four-momentum vectors are recombined to an effective electron 
four-momentum vector if their separation $\Delta R_{e\gamma}$
in the azimuthal angle--pseudo-rapidity plane is smaller than
a critical value $R_c$. If the lepton and photon four-momentum vectors 
are not resolved in the collinear region, the 
collinear singularities from the hard photon contribution can be 
extracted as described in Sec.~IIA, and the integration
over the momentum fraction $z$ in Eq.~(\ref{eq:thirteen}) can be 
performed analytically. The parameter
$\delta_{\theta}$ has to be chosen sufficiently small to ensure that
$\Delta R_{e\gamma}<R_c$ over the entire region where the analytic 
integration
is carried out. For small $R_c$ one finds that $\delta_\theta$ has to be
less than
\begin{equation} 
\delta_{\theta}^{\rm max} \approx {R_c^2 \over 2 \cosh^2(\eta^{\rm
max}(e))}~,
\end{equation}
where $\eta^{\rm max}(e)$ is the maximum allowed pseudo-rapidity of the
electron.

The procedure described above is part of the electron identification
process used by the D\O\ collaboration~\cite{D0Wmass}. The CDF 
collaboration uses a slightly
different method where the electron and photon four-momentum vectors are
combined if both particles traverse the same calorimeter cell. This
modifies the expression for $\delta_\theta^{\rm max}$.

Once the integration over $z$ has been performed analytically, the 
collinear
contribution $d\hat\sigma_{coll.}^{final}$ and the QED-like
contribution $d\hat \sigma^{(0)} F_{Q\!E\!D}^{final}$ can be combined to
cancel the mass singular logarithms explicitly, and one obtains for the
final state $2\to 2$ contribution
\begin{eqnarray}\label{eq:twentyfour}
d\tilde\sigma^{final}_{2\rightarrow 2} &=& \sum_{i,i'} \int dx_1 dx_2 
\; [q_i(x_1,Q^2) \; \overline{q}_{i'}(x_2,Q^2) \; d\hat \sigma^{(0)} 
+(1\leftrightarrow 2) ] \nonumber\\
&\times & \frac{\alpha}{\pi}
\left\{-\ln\delta_s+\frac{11}{4}-\frac{\pi^2}{8}
-\left(\ln\delta_s+
\frac{3}{4}\right)\ln\left(\frac{\delta_{\theta}}{2}\right)
\right\}\; .
\end{eqnarray}

As described in Ref.~\cite{dw}, a part of the photonic interference terms
together with the IR finite parts of the box diagrams in Fig.~\ref{fig:one} 
can be absorbed into the modified weak contributions
$\tilde F_{weak}^{initial,final}$ introduced earlier in this
Section. The $2\rightarrow 2$ interference contribution
is then given by 
\begin{eqnarray}\label{eq:twentyseven}
d\sigma^{interf.}_{2\rightarrow 2} &=& \sum_{i,i'} \int dx_1 dx_2 
\; q_i(x_1,Q^2) \; \overline{q}_{i'}(x_2,Q^2) \; d\hat \sigma^{(0)} 
 \nonumber\\[2.mm]
&\times & \frac{1}{2}\,\beta_{int.}(\hat s, \hat t, \hat u)\;
\ln\left(\frac{\delta_s^2 M_W^4}{(\hat s-M_W^2-\hat s \delta_s)^2
+M_W^2 \Gamma_W^2}\right) +(1\leftrightarrow 2)~, 
\end{eqnarray}
where $\Gamma_W$ is the $W$ width, and 
\begin{equation}
\beta_{int.}(\hat s, \hat t, \hat u)=\frac{\alpha}{\pi}\left[Q_i 
\ln\left(\frac{\hat u^2}{\hat s^2}\right)-Q_{i'}
\ln\left(\frac{\hat t^2}{\hat s^2}\right)+2\right] \; .
\end{equation}
For $\hat s=M_W^2$, the $2\rightarrow 2$ interference contribution
is completely cancelled by the hard photon contribution
$d\hat\sigma_{2\rightarrow 3}^{interf.}$ when the total inclusive cross
section is calculated. Note that the $2\to 2$ interference contribution
exhibits only soft singularities.

The complete ${\cal O}(\alpha^3)$ cross section for $p\,p\hskip-7pt 
\hbox{$^{^{(\!-\!)}}$} \rightarrow W(\gamma)\rightarrow\ell\nu(\gamma)$
can now be expressed as
\begin{eqnarray}\label{eq:twentyeight}
d\sigma^{(0+1)} & = & \sum_{i,i'} \int dx_1 dx_2
\; [q_i(x_1,Q^2) \; \overline{q}_{i'}(x_2,Q^2) \; d\hat \sigma^{(0)} 
+(1\leftrightarrow 2) ] \nonumber \\[2.mm]
&\times & [1+2 \; {\cal R}e \; (\tilde F_{weak}^{initial}+
\tilde F_{weak}^{final})(M_W^2)] 
\nonumber \\[2.mm]
& + & d\sigma^{initial}_{2\rightarrow 2}
+d \sigma_{2\rightarrow 3}^{initial,finite} + 
d\sigma^{interf}_{2\rightarrow 2}
+d \sigma_{2\rightarrow 3}^{interf} +d\sigma^{final}\; ,
\end{eqnarray}
with
\begin{equation}
d\sigma^{final}=d\sigma^{final}_{2\to 2} + d\sigma^{final}_{2\to 3}
\end{equation}
if the integration over $z$ is performed numerically, and
\begin{equation}
d\sigma^{final}=d\tilde\sigma^{final}_{2\to 2} + 
d\sigma^{final,finite}_{2\to 3}
\label{eq:tilde}
\end{equation}
if the $z$-integration is done analytically. Here, 
$d\sigma_{2\rightarrow 3}^{a,finite}$ ($a=initial,~final$) are the reduced
$2 \rightarrow 3$ contributions away from the soft and collinear region. 
The hard bremsstrahlung contribution has been compared
numerically with the $p\,p\hskip-7pt \hbox{$^{^{(\!-\!)}}$} \rightarrow
\ell\nu\gamma$ cross section of Ref.~\cite{BZ}. The two calculations
agree to better than $1\%$. 

\begin{figure}
\vskip 12cm
\includegraphics{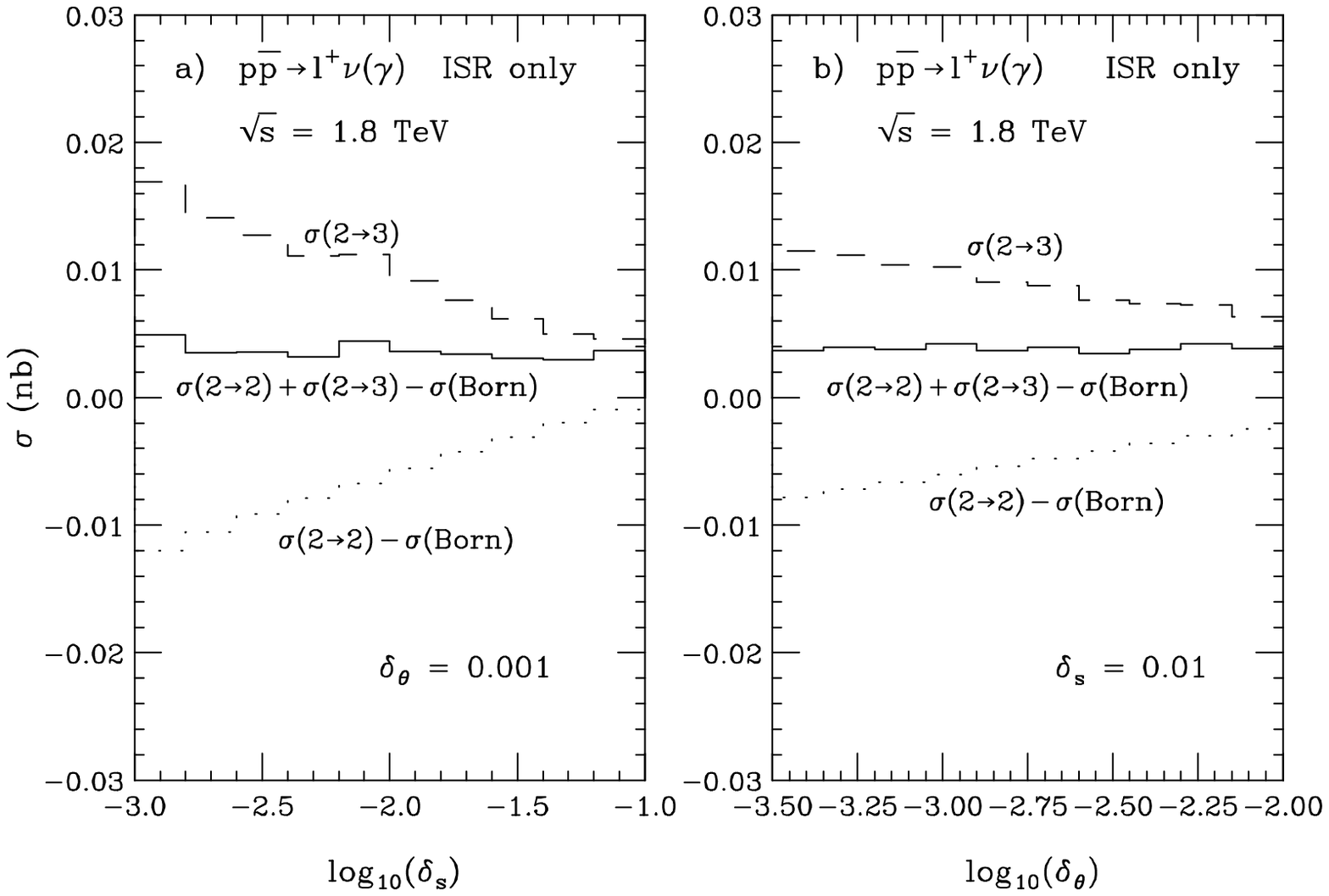}
\vskip -1.cm
\caption{The QED-like initial state corrections to the $p\bar 
p\to\ell^+\nu(\gamma)$, ($\ell=e,\,\mu$) cross 
section for $\protect{\sqrt{s}=1.8}$~TeV as a function of a) $\delta_s$ for
$\delta_\theta=0.001$, and b) $\delta_\theta$ for $\delta_s=0.01$. 
Shown are $\sigma(2\to 2)
-\sigma({\rm Born})$, $\sigma(2\to 3)$, and $\sigma(2\to 2)+\sigma(2\to
3)-\sigma({\rm Born})$. The cuts imposed are listed in
Eqs.~(\protect{\ref{eq:lepcut}}) and~(\protect{\ref{eq:ptmisscut}}). The
energy and momentum resolutions used are described in the text.}
\label{fig:four}
\end{figure}

\begin{figure}
\vskip 12cm
\includegraphics{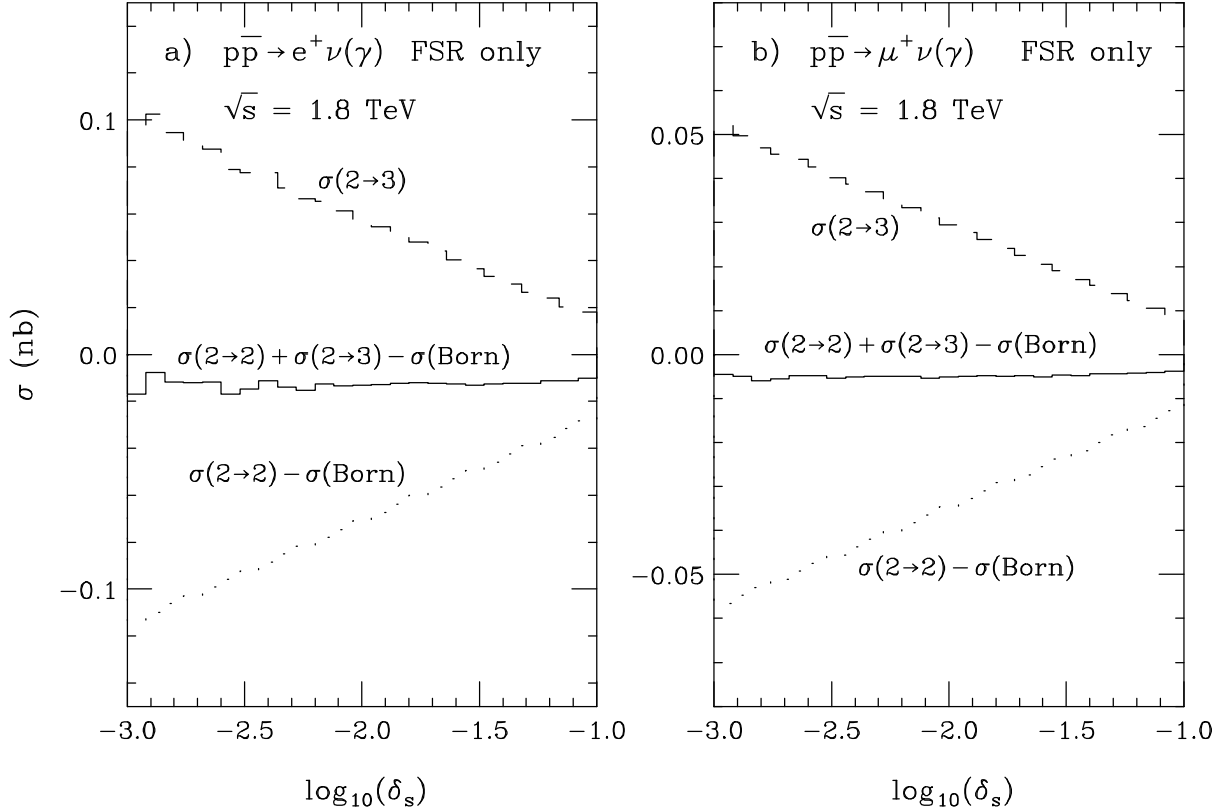}
\vskip -1.cm
\caption{The QED-like final state corrections to the cross section a) 
$\sigma(p\bar p\to e^+\nu(\gamma))$ and b) $\sigma(p\bar 
p\to\mu^+\nu(\gamma))$ for $\protect{\sqrt{s}=1.8}$~TeV as a function
of $\delta_s$. Shown are $\sigma(2\to 2)
-\sigma({\rm Born})$, $\sigma(2\to 3)$, and $\sigma(2\to 2)+\sigma(2\to
3)-\sigma({\rm Born})$. The cuts imposed are listed in
Eqs.~(\protect{\ref{eq:lepcut}}) and~(\protect{\ref{eq:ptmisscut}}). The
energy and momentum resolutions used are described in the text.}
\label{fig:five}
\end{figure}

\begin{figure}
\vskip 12cm
\includegraphics{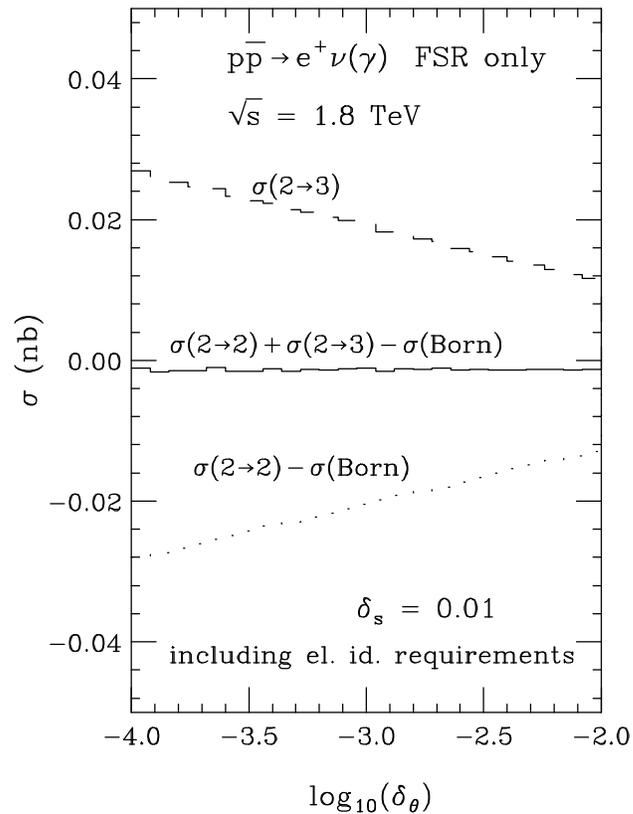}
\vskip -1.cm
\caption{The $p\bar p\to e^+\nu(\gamma)$ cross section for 
$\protect{\sqrt{s}=1.8}$~TeV as a function of $\delta_\theta$ for 
$\delta_s=0.01$ when electron identification requirements are taken into
account and the mass singular terms are canceled analytically. Only 
QED-like final state radiative corrections are included. Shown are 
$\sigma(2\to 2)-\sigma({\rm Born})$, $\sigma(2\to 3)$, and $\sigma(2\to 
2)+\sigma(2\to 3)-\sigma({\rm Born})$. The cuts and lepton
identification requirements imposed are listed in
Eqs.~(\protect{\ref{eq:lepcut}}) and~(\protect{\ref{eq:ptmisscut}}),
and in Table~\protect{\ref{tab:one}}. The energy and momentum resolutions
used are described in the text.}
\label{fig:six}
\end{figure}

The end result of the calculation consists of two sets of weighted
events corresponding to the $2\to 2$ and $2\to 3$ contributions.
Each set depends on the parameters $\delta_s$ and $\delta_\theta$. The 
sum of the two contributions, however, must be independent of $\delta_s$
and $\delta_\theta$, as long as the two parameters are taken small
enough so that the soft photon and the leading pole approximation are
valid. In Figs.~\ref{fig:four} --~\ref{fig:six} we show the different
contributions to the $p\bar 
p\to\ell^+\nu(\gamma)$ cross section at $\sqrt{s}=1.8$~TeV as a function
of the two parameters.
To compute the cross section, we use here and in all subsequent figures 
the MRSA set of parton distribution functions~\cite{MRSA}, and take the
QCD renormalization scale $\mu_{QCD}$ and the QED and QCD factorization 
scales, $M_{QED}$ and $M_{QCD}$, to be 
$\mu^2_{QCD}=Q^2=M_{QED}^2=M_{QCD}^2=M_W^2$. 
The detector acceptance is simulated by imposing the following
transverse momentum ($p_T$) and pseudo-rapidity ($\eta$) cuts:
\begin{equation}
p_T(\ell)>25~{\rm GeV,}\qquad\qquad |\eta(\ell)|<1.2, \qquad\qquad
\ell=e,\,\mu ,
\label{eq:lepcut}
\end{equation}
\begin{equation}
p\llap/_T>25~{\rm GeV.}
\label{eq:ptmisscut}
\end{equation}
These cuts approximately model the acceptance cuts used by the CDF and D\O\
collaborations in their $W$ mass analyses~\cite{cdfwmass,D0Wmass}.
Uncertainties in the energy and momentum measurements of the charged leptons 
in the detector are simulated in the calculation by Gaussian smearing 
of the particle four-momentum vector with standard deviation $\sigma$
which depends on the particle type and the detector. The numerical results 
presented here were calculated using $\sigma$ values based on the 
specifications for the upgraded Run~II D\O\ detector~\cite{d0upgr}. The 
results obtained using the target specifications for
the CDF~II detector~\cite{cdfii} are similar.
The SM parameters used in our numerical simulations are $M_W=80.3$~GeV,
$M_Z=91.187$~GeV, $\alpha=\alpha(0)=1/137.036$, $G_\mu=1.166\times 
10^{-5}~{\rm GeV}^{-2}$, $\Gamma_W=2.1$~GeV, and $m_{top}=175$~GeV. 
These values are consistent with recent measurements at LEP, SLC and the 
Tevatron~\cite{holliksm}. 

Figure~\ref{fig:four} displays the QED-like initial state (ISR) 
corrections to the cross section as a function
of $\delta_s$ (Fig.~\ref{fig:four}a) and $\delta_\theta$ 
(Fig.~\ref{fig:four}b). In order to exhibit the independence of the
cross section from the parameters $\delta_s$ and $\delta_\theta$ more
clearly, we have not included the Born cross section in the $2\to 2$
contribution here as well as in Figs.~\ref{fig:five} and~\ref{fig:six}.
The QED-like ISR corrections to the cross section for electron and muon 
final states are virtually identical. While the separate $2\to 2$ and 
$2\to 3$ ${\cal O}(\alpha)$ 
contributions vary strongly with $\delta_s$ and $\delta_\theta$, the sum is
independent of the two parameters within the accuracy of the Monte Carlo
integration. 

In Fig.~\ref{fig:five}, we show the QED-like final state (FSR) 
corrections to
the $p\bar p\to\ell^+\nu(\gamma)$ cross section as a function of 
$\delta_s$ for electron and muon final states. 
Radiation of photons collinear with one of the leptons gives rise to
terms proportional to $[\ln(\hat s/m^2_\ell)-2]\ln(\delta_s)$ (see 
Eq.~(\ref{eq:twentythree})) in both the $2\to 2$ and $2\to 3$ contributions.
As demonstrated in Fig.~\ref{fig:five}, these terms cancel
and the total cross section is independent of $\delta_s$. Due to the
smaller mass of the electron, the variation of the $2\to 2$ and $2\to 3$
contributions with $\delta_s$ is more pronounced in the electron case.

In Figs.~\ref{fig:four} and~\ref{fig:five}, we have not taken into 
account realistic lepton
identification requirements, {\it i.e.} we have assumed that photons and 
leptons with arbitrary small opening angles can be discriminated. 
In a more realistic simulation, in addition to the
lepton $p_T$, $p\llap/_T$ and pseudo-rapidity cuts, one imposes
requirements on the separation of the charged lepton and the
photon. These requirements differ slightly for the CDF and D\O\
detectors. In the following we adopt lepton identification criteria
which are motivated by the D\O\ $W$ mass~\cite{D0Wmass} and $W$ cross 
section~\cite{D0Wcross} analyses; the numerical results obtained using 
the requirements imposed in the CDF $W$ mass
analysis~\cite{cdfwmass} are similar. In order to study their impact on 
the size of the EW radiative corrections, we will perform simulations 
both with and without the lepton identification requirements taken into 
account.
 
We shall use the following lepton identification requirements. For
electrons, we require that the electron and photon 
momentum four-vectors are combined into an 
effective electron momentum four-vector if $\Delta R_{e\gamma}<0.2$. For
$0.2<\Delta R_{e\gamma}<0.4$ events are rejected if
$E_{\gamma}>0.15 \; E_e$. Here $E_\gamma$ ($E_e$) is the energy of the
photon (electron) in the laboratory frame. For events with $0.2<\Delta 
R_{e\gamma}<0.3$ and 
$E_{\gamma}<0.15 \; E_e$, the electron and photon momentum four-vectors 
are again combined. Muons are identified by hits in the muon chambers 
and the requirement that the associated track is consistent with a 
minimum ionizing particle. This limits the photon energy for small 
muon -- photon opening angles. For muons, we therefore require the energy 
of the photon to be $E_{\gamma}<2$~GeV for $\Delta R_{\mu\gamma}<0.2$, and 
$E_{\gamma}<6$~GeV for $0.2<\Delta R_{\mu\gamma}<0.6$. For future 
reference, we summarize the lepton identification requirements in 
Table~\ref{tab:one}.

As noted before, 
when the electron and photon momentum four-vectors are combined, it is
possible to analytically cancel the mass singular terms in the QED-like
$2\to 2$ final state corrections. In this case, the QED-like final state
$2\to 2$ and
$2\to 3$ contributions depend on the collinear cutoff parameter
$\delta_\theta$. Figure~\ref{fig:six} displays the QED-like final state
contribution to the $p\bar p\to
e^+\nu(\gamma)$ cross section as a function of $\delta_\theta$ when the
electron identification requirements described above are taken into
account. While the $2\to 2$ and $2\to 3$ contributions both exhibit a
considerable dependence on $\delta_\theta$, their sum is independent of
the parameter. 

Similar to the QED-like initial and final state corrections, one can 
show that the sum of the $2\to 2$ and $2\to 3$ contributions of the 
QED-like initial -- final state interference terms is independent of
$\delta_s$. The interference terms are typically of the same size as the
initial state corrections. The modified weak contributions to the ${\cal
O}(\alpha^3)$ cross section are trivially independent of $\delta_s$ and
$\delta_\theta$. In the following, these parameters will be fixed to 
$\delta_s = 10^{-2}$ and $\delta_\theta = 10^{-3}$. 

As stated before, we take the QCD renormalization scale $\mu_{QCD}$ and 
the QED and QCD factorization scales, $M_{QED}$ and $M_{QCD}$, to be equal,
$\mu_{QCD}=M_{QED}=M_{QCD}=Q$.
The missing QED corrections to the PDF's create a dependence of the 
${\cal O}(\alpha)$ initial state corrections on the scale
$Q$ which is stronger than that of the lowest order calculation.
On the other hand, final state and initial -- final state interference 
terms depend on $Q$ only through the PDF's. 

\section{Phenomenological Results}

We shall now discuss the phenomenological implications of the ${\cal
O}(\alpha)$ electroweak corrections to $W$ production at the Tevatron
($p\bar p$ collisions at $\sqrt{s} = 1.8$~TeV). We first discuss the
impact of electroweak corrections on observables used to measure $M_W$:
the transverse mass distribution, the $p_T(\ell)$ distribution, and the
$W$ to $Z$ transverse mass ratio. 
We then consider the $W$ production cross section, the $W$ to $Z$
cross section ratio and the charge asymmetry of leptons in the $W$
decay. Unless stated otherwise, we take into account the cuts
of Eqs.~(\ref{eq:lepcut}) and~(\ref{eq:ptmisscut}) and effects from 
energy and momentum measurement uncertainties in the detector. We state 
explicitly when the lepton identification requirements listed in
Table~\ref{tab:one} are included.

\subsection{Electroweak Corrections to the $M_T$ and $p_T(\ell)$ 
Distributions, and the $W$ to $Z$ Transverse Mass Ratio}

Since the detectors at the Fermilab Tevatron collider cannot directly 
detect the neutrinos produced in the leptonic $W$ boson decays,
$W\to\ell\nu$, and cannot measure the
longitudinal component of the recoil momentum, there is insufficient
information to reconstruct the invariant mass of the $W$ boson. 
Instead, the transverse mass distribution of the final state
lepton pair, or the transverse momentum distribution of the charged
lepton are used~\cite{D0Wmass} to extract $M_W$. The transverse mass is
defined by
\begin{equation}
M_T=\sqrt{2p_T(\ell)p_T(\nu)(1-\cos\phi^{\ell\nu})}~,
\label{eq:mt}
\end{equation}
where $p_T(\ell)$ and $p_T(\nu)$ are the transverse momentum of the 
lepton and the neutrino, and $\phi^{\ell\nu}$ is the angle between the 
charged lepton and the neutrino in the transverse plane. The neutrino 
transverse momentum is identified with the missing transverse momentum,
$p\llap/_T$, in the event. Recently, it has been pointed out that the
ratio of $W$ to $Z$ observables can also be used to measure the $W$
mass~\cite{GK}. This method has been applied to the $W$ to
$Z$ transverse mass ratio by the D\O\ collaboration~\cite{srini}. 
The advantages and disadvantages of the observables used to extract
$M_W$ are discussed in Ref.~\cite{BD}. 

\begin{figure}
\vskip 12cm
\includegraphics{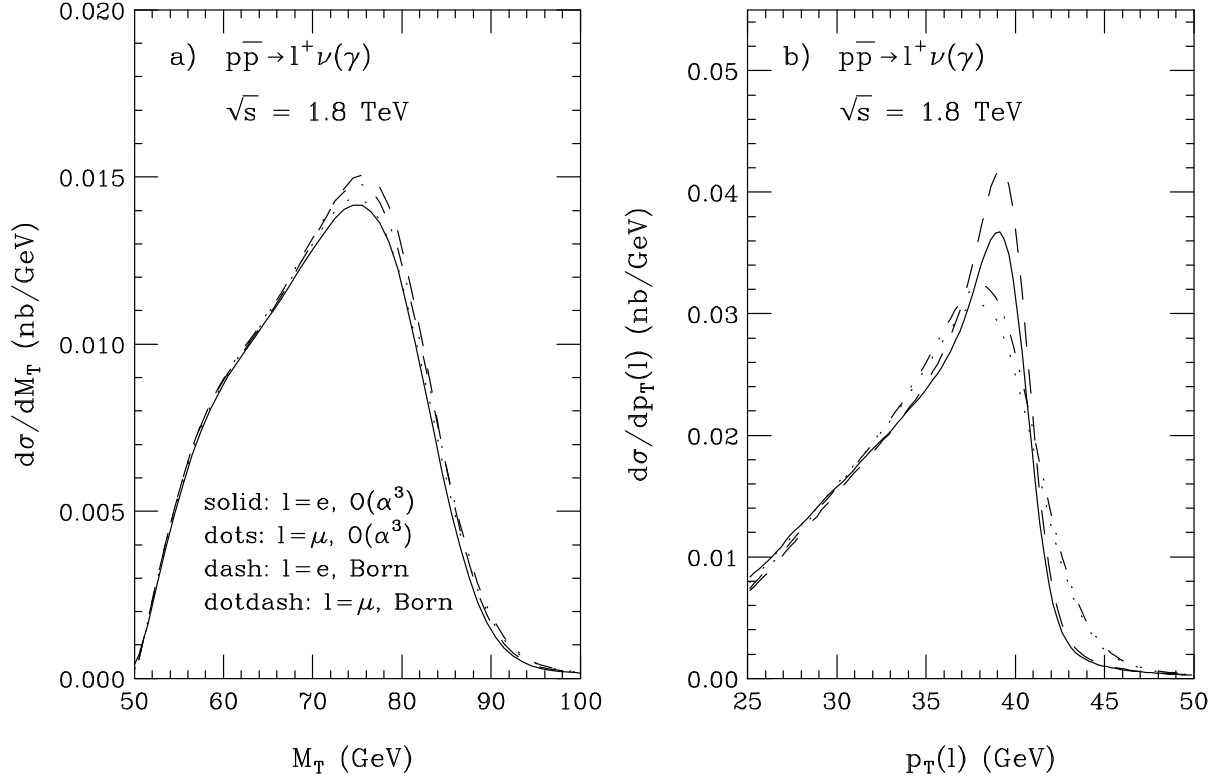}
\vskip -1.cm
\caption{Differential cross sections for $p\bar p\to
\ell^+\nu(\gamma)$ at $\protect{\sqrt{s}=1.8}$~TeV. Shown in part a) is
the transverse mass distribution. Part b) displays the lepton transverse
momentum spectrum. The solid (dotted) lines show the distributions for
electron (muon) final states including ${\cal O}(\alpha)$ electroweak
corrections. The dashed (dot-dashed) line gives the $e^+\nu$
($\mu^+\nu$) Born cross section. The cuts imposed are listed in
Eqs.~(\protect{\ref{eq:lepcut}}) and~(\protect{\ref{eq:ptmisscut}}). The
energy and momentum resolutions used are described in Sec.~IIC. The 
lepton identification requirements of Table~\protect{\ref{tab:one}} are 
not taken into account here. }
\label{fig:seven}
\end{figure}

The ${\cal O}(\alpha^3)$ $M_T$ distribution for $e^+\nu(\gamma)$ (solid)
and $\mu^+\nu(\gamma)$ (dots) production is shown in
Fig.~\ref{fig:seven}a together with the lowest order predictions (dashed
and dot-dashed curves). In Fig.~\ref{fig:seven}b, we show the 
${\cal O}(\alpha^3)$ and Born $p_T(\ell)$ spectrum. The flavor specific 
lepton identification requirements of Table~\ref{tab:one} are not 
taken into account here. Electroweak corrections decrease
the cross section at the peak of the $M_T$ ($p_T(\ell)$) distribution by 
about $12\%$ ($17\%$) in the electron, and by about $6\%$ ($7\%$) in the
muon case. Photon radiation from the charged
lepton lowers the $\ell\nu$ invariant mass. Events from the Jacobian
peak regions in the $M_T$ and $p_T(\ell)$ distributions therefore are 
shifted on average to lower values of the transverse mass and transverse
momentum. Due to the $\ln(\hat s/m_\ell^2)$ term, the effect of the 
corrections is larger in the electron case. The Jacobian 
peak of the $p_T$ distribution is broader and less pronounced in the
muon case, due to the energy and momentum resolution which is
significantly worse for muons than for electrons. In the $M_T$ 
distribution, the effect of the $p\llap/_T$ resolution dominates, and 
the difference in the distribution between electrons and muons is small.
The $M_T$ and $p_T(\ell)$ distributions for $\ell^-\nu(\gamma)$
production are identical to those for the $\ell^+\nu(\gamma)$ channel in
$p\bar p$ collisions. In the remainder of this subsection we therefore
only consider the $\ell^+\nu(\gamma)$ final state. 

\begin{figure}
\vskip 12cm
\includegraphics{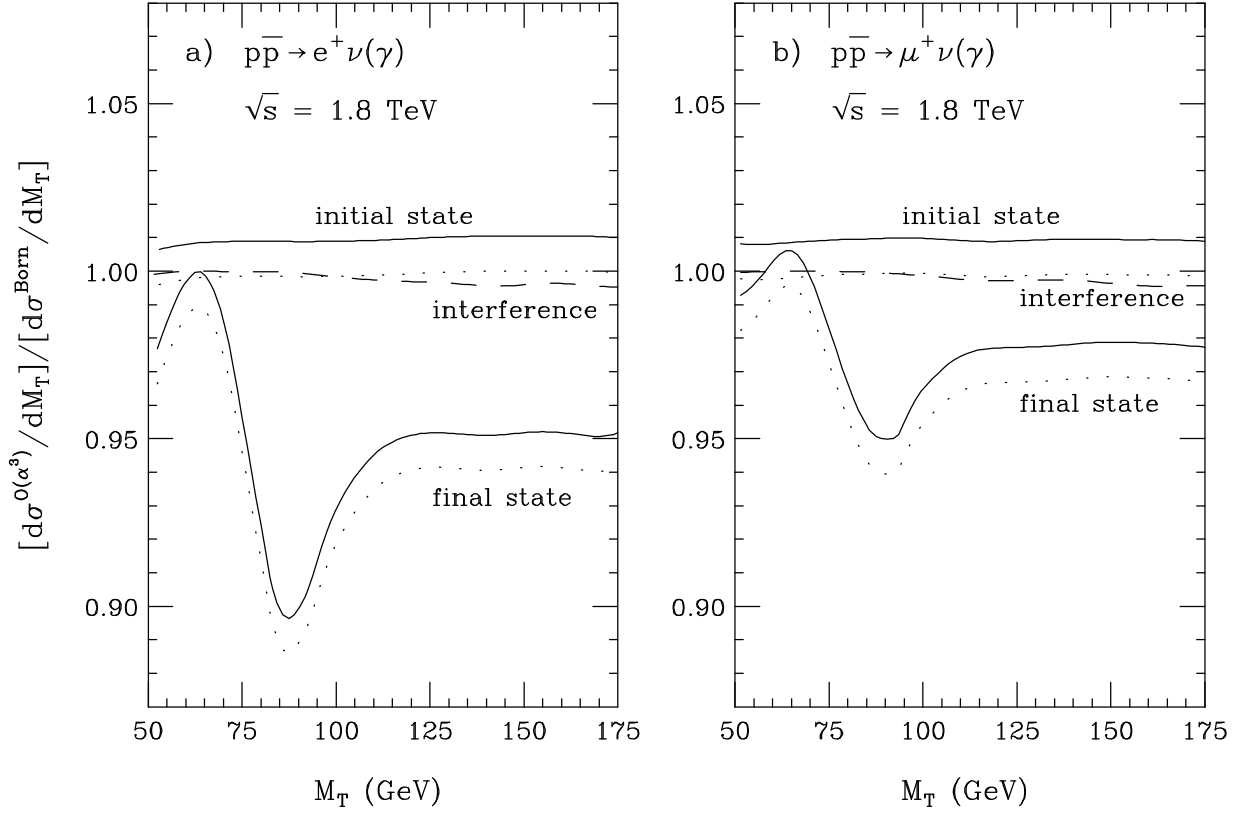}
\vskip -1.cm
\caption{Ratio of the ${\cal O}(\alpha^3)$ and lowest order 
cross sections as a function of the transverse mass for a) $p\bar p\to
e^+\nu(\gamma)$ and b) $p\bar p\to \mu^+\nu(\gamma)$ at
$\protect{\sqrt{s}=1.8}$~TeV for various individual contributions. 
The upper (lower) solid lines show the result for the QED-like initial 
(final) state corrections. The upper (lower) dotted lines give the cross
section ratios if
both the QED-like and modified weak initial (final) state corrections are 
included. The dashed lines display the result if only the 
initial -- final state interference contributions are included.
The cuts imposed are listed in Eqs.~(\protect{\ref{eq:lepcut}}) 
and~(\protect{\ref{eq:ptmisscut}}). The energy and momentum resolutions
used are described in Sec.~IIC. The lepton identification requirements of
Table~\protect{\ref{tab:one}} are not taken into account here. }
\label{fig:eight}
\end{figure}

The various individual contributions to the
EW ${\cal O}(\alpha)$ corrections on the $M_T$ distribution are shown
in Fig.~\ref{fig:eight}. The initial state QED-like 
contribution uniformly increases  the cross section by about 1\% for 
electron (Fig.~\ref{fig:eight}a) and muon (Fig.~\ref{fig:eight}b) final 
states. It is largely canceled by the modified weak initial state 
contribution. The interference contribution is very small. It decreases
the cross section by about $0.01\%$ for transverse masses below $M_W$, and
by up to $0.5\%$ for $M_T>M_W$. The final state QED-like contribution 
significantly changes 
the shape of the transverse mass distribution and reaches its maximum
effect in the region of the Jacobian peak, $M_T\approx M_W$. As for 
the initial state,
the modified weak final state contribution reduces the cross section by 
about $1\%$, and has no effect on the shape of the transverse 
mass distribution. For $M_T>125$~GeV, the QED-like final state
corrections uniformly reduce the differential cross section by about $5\%$
in the electron case, and by about $2\%$ in the muon case. Without
taking the lepton identification requirements of Table~\ref{tab:one}
into account, the full ${\cal O}(\alpha)$ electroweak radiative
corrections to the $M_T$ distribution are very well approximated by the
sum of the QED-like and modified weak final state corrections. 

It should be noted that the differential cross section ratio shown in
Fig.~\ref{fig:eight} becomes ill defined in the threshold region
$M_T\approx p_T^{cut}(\ell)+p\llap/_T^{cut}$, where $p_T^{cut}(\ell)$
and $p\llap/_T^{cut}$ are the charged lepton $p_T$ and the missing 
transverse momentum threshold. For 
$M_T\leq p_T^{cut}(\ell)+p\llap/_T^{cut}$, the 
Born cross section vanishes, and the cross section ratio is
infinite. The ${\cal O}(\alpha^3)$
cross section is small, but non-zero, in this region. The largest
contribution to the cross section for $M_T\leq
p_T^{cut}(\ell)+p\llap/_T^{cut}$ originates from initial state radiation
configurations, where the lepton and the neutrino have a small relative 
opening angle and are balanced by a high $p_T$ photon in the opposite 
hemisphere. Close to the
threshold, $M_T\approx p_T^{cut}(\ell)+p\llap/_T^{cut}$, large logarithmic
corrections are present, and for an accurate prediction in this region 
those corrections need to be resummed. The results of 
Fig.~\ref{fig:eight} in this region should therefore be interpreted with
caution. 

\begin{figure}
\vskip 12cm
\includegraphics{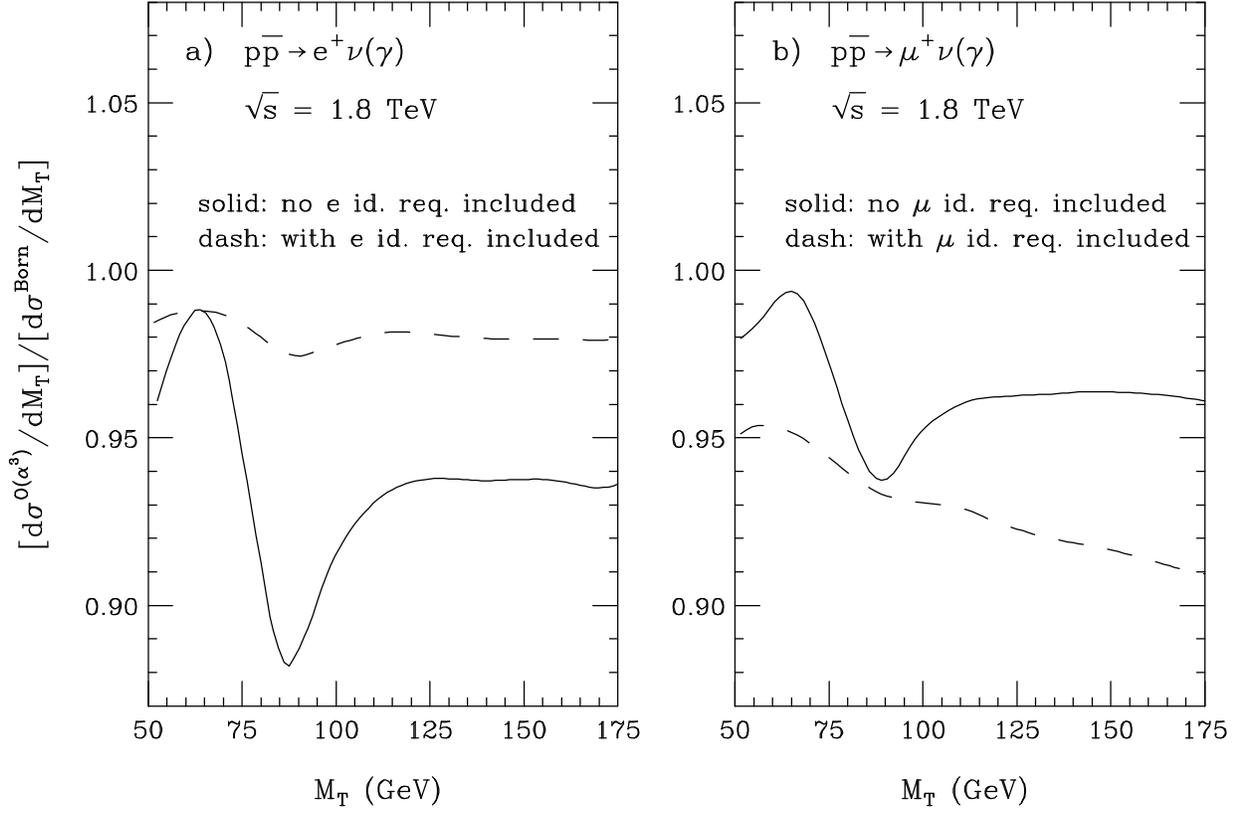}
\vskip -1.cm
\caption{Ratio of the full ${\cal O}(\alpha^3)$ and lowest order differential
cross sections as a function of the transverse mass for a) $p\bar p\to
e^+\nu(\gamma)$ and b) $p\bar p\to \mu^+\nu(\gamma)$ at
$\protect{\sqrt{s}=1.8}$~TeV. The dashed (solid) lines show the
result with (without) the lepton identification requirements of 
Table~\protect{\ref{tab:one}} taken into account. The cuts imposed are 
listed in Eqs.~(\protect{\ref{eq:lepcut}})
and~(\protect{\ref{eq:ptmisscut}}). The energy and momentum resolutions
used are described in Sec.~IIC.} 
\label{fig:nine}
\end{figure}

The ratio of the full ${\cal O}(\alpha^3)$ and the Born cross section as
a function of the transverse mass is shown in Fig.~\ref{fig:nine}. The
solid (dashed) lines show the cross section ratio without (with) the 
lepton identification requirements included. Recombining the electron 
and photon
four-momentum vectors for $\Delta R_{e\gamma}<0.2$ eliminates the 
mass singular logarithmic terms and strongly reduces the size of the
QED-like final state corrections (see Fig.~\ref{fig:nine}a). These 
corrections are now of the same size as the initial state QED-like and 
the modified weak corrections. However, with the total ${\cal 
O}(\alpha)$ EW corrections varying between $1\%$ and $2\%$, the shape 
change of the $M_T$ distribution caused by the final state corrections 
is still significant. For muon final
states (see Fig.~\ref{fig:nine}b), the cut on the energy of the photon 
reduces the hard photon part of the ${\cal O}(\alpha^3)$
$\mu\nu(\gamma)$ cross section. In this case, the mass singular terms
survive and the corrections become larger
over the entire range of $M_T$ considered. Before lepton
identification requirements are taken into account, the change in the
shape of the $M_T$ distribution due to the QED-like final state
corrections is more pronounced in the electron channel. Once these 
requirements are included, the shape change is stronger in the muon case. 

Results qualitatively similar to those shown in Figs.~\ref{fig:eight}
and~\ref{fig:nine} are obtained for the transverse momentum distribution 
of the charged lepton. 

As we have seen, final state bremsstrahlung has a non-negligible effect
on the $M_T$ and $p_T(\ell)$ distribution in the Jacobian peak
region. As is well known, electroweak corrections must be included when the
$W$ boson mass is extracted from data, otherwise the mass is 
shifted to a lower value. In the approximate treatment of
the electroweak corrections used so far by the Tevatron experiments,
only final state QED corrections are taken into account; initial state,
interference, and weak correction terms are ignored. Furthermore, the 
effect of the final state soft and virtual photonic
corrections is estimated from the inclusive ${\cal O}(\alpha^2)$ 
$W\to\ell\nu(\gamma)$ width~\cite{Albert} and the hard photon 
bremsstrahlung contribution~\cite{BK,RGW}. When detector effects are
included, the approximate calculation leads to a shift of about
$-50$~MeV in the electron case, and approximately $-160$~MeV in the muon
case~\cite{cdfwmass,D0Wmass}. 

\begin{figure}
\vskip 12cm
\includegraphics{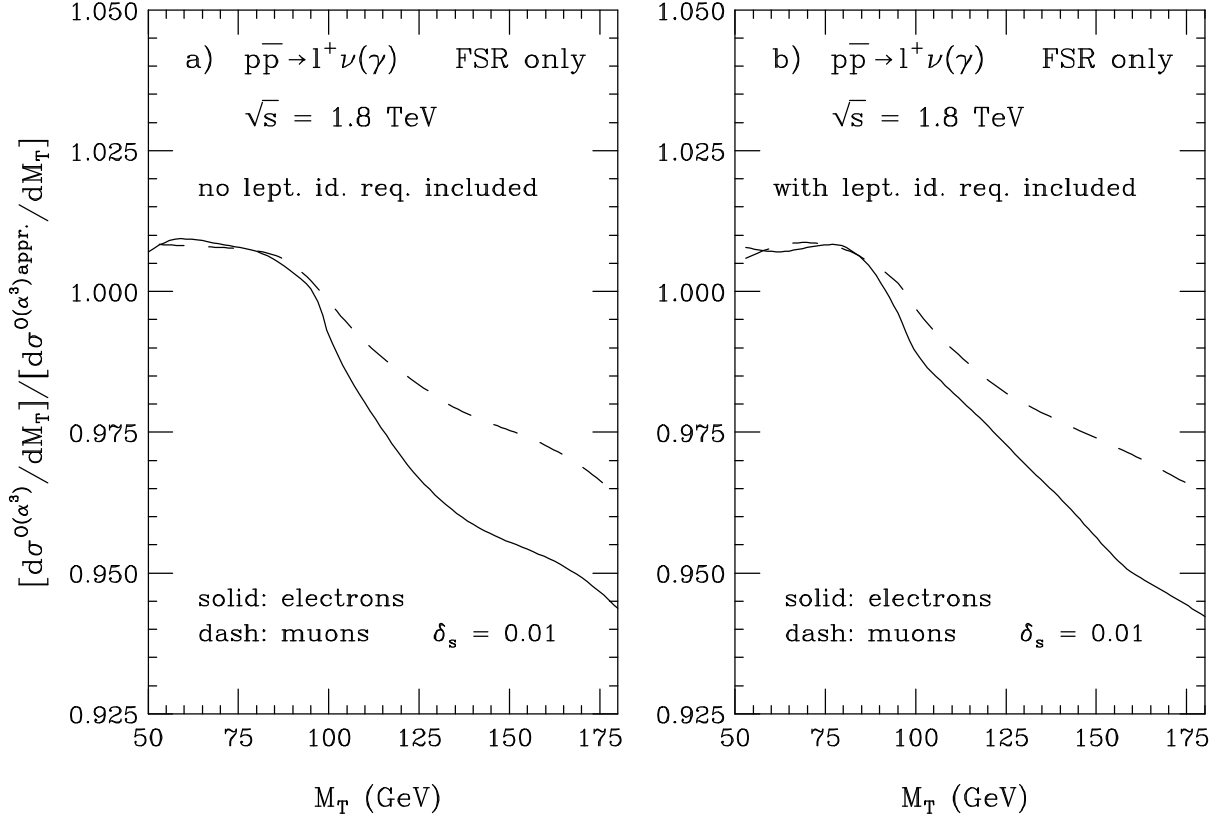}
\vskip -1.cm
\caption{Ratio of the $M_T$ distributions obtained with the QED-like
final state correction part of our calculation to the one obtained using
the approximation of Refs.~[\ref{ber}] and~[\ref{wag}] for
$p\bar p\to\ell^+\nu(\gamma)$ at $\protect{\sqrt{s}=1.8}$~TeV a) without
and b) with lepton identification requirements (see
Table~\protect{\ref{tab:one}}) taken into account. The solid and dashed
lines give the results for electron and muon final states,
respectively. The approximate NLO transverse mass distribution
does depend on $\delta_s$ (see Eq.~(\protect{\ref{eq:twentyfive}})) 
which is taken to be $\delta_s=0.01$.
The cuts imposed are listed in Eqs.~(\protect{\ref{eq:lepcut}})
and~(\protect{\ref{eq:ptmisscut}}). The energy and momentum resolutions
used are described in Sec.~IIC.} 
\label{fig:ten}
\end{figure}

Initial state and 
interference contributions do not change the shape of the $M_T$
distribution significantly (see Fig.~\ref{fig:eight}) and therefore have
little effect on the
extracted mass. However, correctly incorporating the final state 
virtual and soft photonic corrections results in a non-negligible 
modification of the shape of the transverse mass distribution. 
This is demonstrated in
Fig.~\ref{fig:ten}, which shows the ratio of the $M_T$ distribution
obtained with the QED-like final state correction part of our
calculation to the one obtained using the approximation of
Refs.~\cite{BK} and~\cite{RGW}. The dependence of the ratio on $M_T$ is
described by Eq.~(\ref{eq:twentyfive}). For $M_T<M_W$, most events originate
from the region $\hat s\approx M_W^2$, due to the Breit-Wigner
resonance. Consequently, there is little dependence of the cross section
ratio on $M_T$ in this region. For $M_T>M_W$, the steeply falling cross
section in the tail of the Breit-Wigner resonance favors events with
$\hat s\approx M_T^2$. In this region the term proportional to
$\ln(\hat s/M_W^2)$ in Eq.~(\ref{eq:twentyfive})
causes a change in the shape of the transverse
mass distribution. $\Delta\hat\sigma^{final}$ also contains a term which is 
proportional to $\ln(m^2_\ell/M_W^2)$ (see Eq.~(\ref{eq:twentyfive})). The
shape change in the $M_T$ distribution thus is more pronounced in the
electron case. Lepton identification requirements have a small effect on
the cross section ratio (see Fig.~\ref{fig:ten}). Note
that the approximate NLO cross section, and thus the cross section ratio
shown in Fig.~\ref{fig:ten}, does depend explicitly on the cutoff
$\delta_s$ whereas the ${\cal O}(\alpha^3)$ cross section resulting from
our calculation does not. While the dependence on $\delta_s$ is 
very small for $M_T<M_W$, it is quite pronounced for transverse masses 
above $M_W$. 

The difference in the line shape of the $M_T$ distribution between the
complete ${\cal O}(\alpha^3)$ calculation and the approximation 
used so far occurs in a region which is important for both the
determination of the $W$ mass, and the direct measurement of the $W$
width. The precision which can be achieved in a measurement of $M_W$
using the transverse mass distribution strongly depends on how steeply
the $M_T$ distribution falls in the region $M_T\approx M_W$ (see
Fig.~\ref{fig:seven}). In the region of large transverse masses, $M_T> 
100\dots 110$~GeV, the shape of the $M_T$ distribution is sensitive to 
the $W$ width~\cite{cdfwwidth}. Any change in the theoretical prediction
of the line shape thus directly influences the $W$ mass and width
measurements. From a maximum likelihood analysis similar to that carried
out in Ref.~\cite{BKS} for $Z$ production, the shift in the measured $W$
mass due to the correct
treatment of the final state virtual and soft photonic corrections is
found to be $\Delta M_W\approx {\cal O}(10~{\rm MeV})$. This shift is 
much smaller than the present uncertainty for
$M_W$ from hadron collider experiments~\cite{cdfwmass,D0Wmass}. However,
for future precision experiments, a difference of ${\cal O}(10~{\rm
MeV})$ in the extracted value of $M_W$ can no longer be ignored, and the
complete ${\cal O}(\alpha^3)$ calculation should be used. 

At high luminosities, the transverse mass ratio of $W$ to $Z$ bosons
offers advantages in determining the $W$ mass~\cite{GK,srini} over the
$M_T$ and $p_T(\ell)$ distributions. The
transverse mass ratio of $W$ and $Z$ bosons is defined as
\begin{equation}
R_{M_T}(X_{M_T})={A_W(X^W_{M_T}=X_{M_T})\over A_Z(X^Z_{M_T}=X_{M_T})}~,
\end{equation}
where $A_V$ ($V=W,\,Z$) is the differential cross section
\begin{equation}
A_V(X^V_{M_T})={d\sigma_V\over dX^V_{M_T}}
\end{equation}
with respect to the scaled transverse mass,
\begin{equation}
X^V_{M_T}={M^V_T\over M_V}~.
\end{equation}
The transverse mass of the lepton pair in $Z$ boson events is defined in
complete analogy to Eq.~(\ref{eq:mt}):
\begin{equation}
M^Z_T=\sqrt{2p_T(\ell^+)p_T(\ell^-)(1-\cos\phi)}~,
\end{equation}
where $\phi$ is the angle between the two charged leptons in the 
transverse plane.

\begin{figure}
\vskip 12cm
\includegraphics{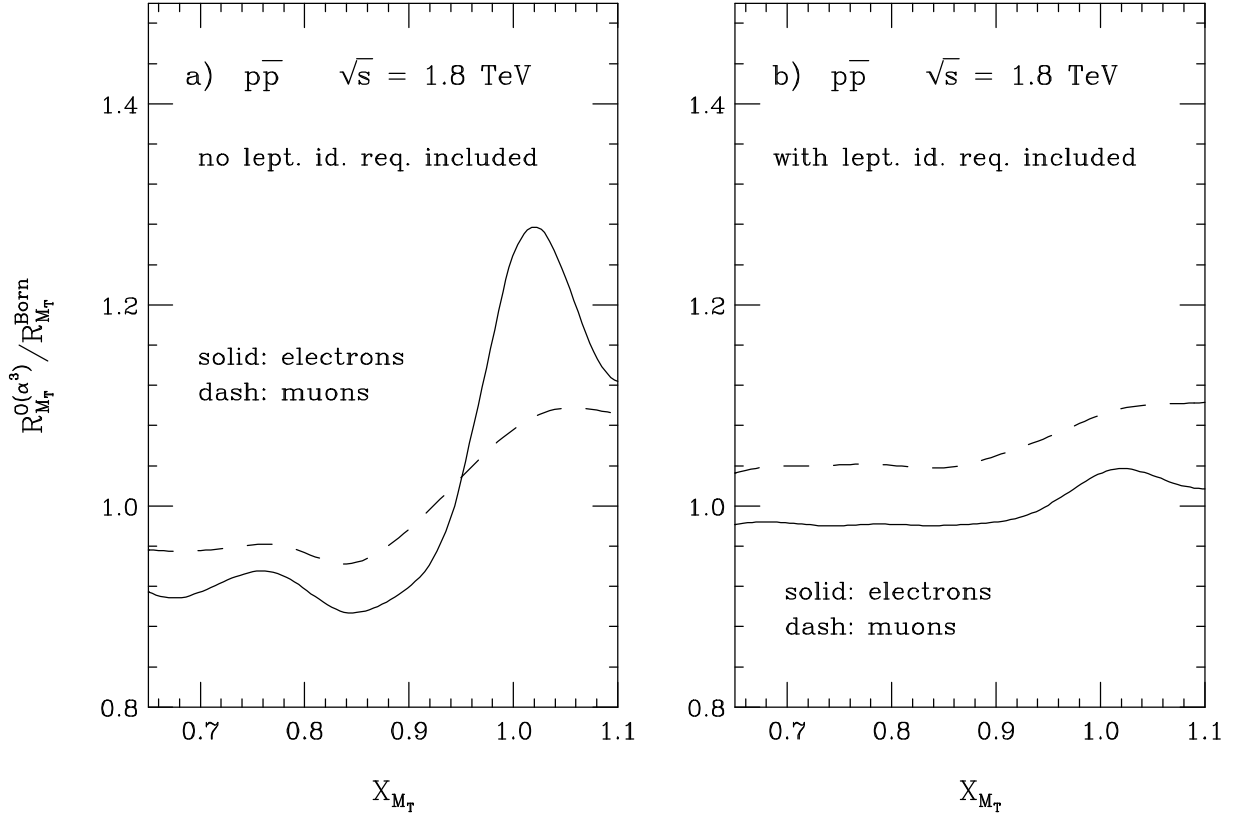}
\vskip -1.cm
\caption{Ratio of the ${\cal O}(\alpha^3)$ and lowest order $W^+$ to
$Z$ transverse mass ratio as a function of the scaled transverse mass,
$X_{M_T}$, at $\protect{\sqrt{s}=1.8}$~TeV. The solid 
(dashed) lines show the result for the electron (muon) final state. The
ratio without and with lepton identification requirements (see
Table~\protect{\ref{tab:one}}) taken into account is shown in part a)
and part b) of the figure, respectively. The cuts imposed are listed in 
Eqs.~(\protect{\ref{eq:lepcut}}) and~(\protect{\ref{eq:ptmisscut}}). The
energy and momentum resolutions used are described in Sec.~IIC. For
$p\bar p\to\ell^+\ell^-(\gamma)$, we in addition require the di-lepton 
invariant mass to satisfy the constraint $75~{\rm
GeV}<m(\ell^+\ell^-)<105~{\rm GeV}$.} 
\label{fig:eleven}
\end{figure}

The ratio of the ${\cal O}(\alpha^3)$ and the Born $W$ to $Z$ transverse
mass ratio is shown in Fig.~\ref{fig:eleven}. To calculate the ${\cal
O}(\alpha)$ electroweak corrections to $Z$ boson production, we use the
results of Ref.~\cite{BKS}. Note that purely weak corrections are not
included in this calculation. Identical
charged lepton $p_T$ and rapidity cuts are used for $W$ and $Z$
production. In the $Z$ boson case, photon exchange and $\gamma Z$
interference effects are included and an additional cut on the di-lepton 
invariant mass of $75~{\rm GeV}<m(\ell^+\ell^-)<105~{\rm GeV}$ has been
imposed. In $p\bar p\to\ell\nu(\gamma)$ only one of the two leptons can 
emit a photon, whereas both leptons in $p\bar p\to\ell^+\ell^-(\gamma)$ 
can radiate. The ${\cal O}(\alpha)$ corrections are thus significantly
larger in the $Z$ case. As a result, the $W$ to $Z$ transverse
mass ratio is more strongly affected by electroweak radiative
corrections than the $M_T^W$ distribution. Without the lepton 
identification requirements, the ${\cal O}(\alpha)$ 
corrections increase $R_{M_T}$ by about $30\%$ ($10\%$) at the location of 
the Jacobian peak ($X_{M_T}=1$) for electrons (muons). For
$X_{M_T}<0.9$, the 
electroweak corrections reduce the transverse mass ratio by $6-10\%$ in
the electron case, and by $4-6\%$ in the muon case. 

When lepton identification criteria are taken into account, the 
merging of the electron and photon momentum four-vectors for small
$e-\gamma$ opening angles again strongly reduces the size of the ${\cal 
O}(\alpha)$ corrections (see Fig.~\ref{fig:eleven}b). In 
the region of the Jacobian peak, the corrections are reduced to 
$\approx 4\%$, and for $X_{M_T}<0.95$ to about $2\%$ in magnitude. For 
muon final states, the lepton identification requirements reduce the 
hard photon part of the ${\cal O}(\alpha^3)$ cross sections below the
Jacobian peak, but have little effect in the peak region where only few
events contain hard photons (see Fig.~\ref{fig:nine}b). The reduction 
is more pronounced for the
$p\bar p\to\mu^+\mu^-(\gamma)$ than for the $p\bar p\to\mu\nu(\gamma)$ 
cross section. Consequently, the ${\cal O}(\alpha)$ corrections increase
$R_{M_T}$ below the Jacobian peak and leave it almost unchanged in the 
peak region. 

\subsection{Electroweak Corrections to the $W$ Boson Cross Section and 
the $W$ to $Z$ Cross Section Ratio}

In the past, the measurement of the $W$ and $Z$ boson cross sections has
provided a test of perturbative QCD~\cite{D0Wcross,CDFZSIG,CDFZSIGN}.
With the large data set accumulated in the 1994-95 Tevatron collider 
run, the uncertainty associated with the integrated luminosity ($\approx
3.6\%$~\cite{CDFZSIGN}) became a limiting factor in this measurement. This
suggests to use the measured $W$ and $Z$ boson cross sections to 
determine the integrated luminosity in Run~II~\cite{CDFZSIGN,DITT1}. 
The cross section ratio
\begin{equation}
R_{W/Z}={\sigma(p\bar p\to W\to\ell\nu X)\over\sigma(p\bar p\to Z
\to\ell^+\ell^- X)}~,
\end{equation}
together with the theoretical prediction for the ratio of the total $W$
and $Z$ production cross sections, 
$\sigma_W/\sigma_Z=3.36\pm 0.02$~\cite{ADM}, the LEP measurement of 
the branching ratio
$B(Z\to\ell^+\ell^-)$ and the SM prediction for the $W\to\ell\nu$ decay
width, can be used for an indirect determination of
$\Gamma_W$~\cite{D0Wcross,cdfr}. For integrated luminosities smaller
than about $20~{\rm fb}^{-1}$, the $W$ width measurement from $R_{W/Z}$
is expected to yield better results than the direct determination from 
the $M_T$ distribution~\cite{Tev2000}. 

The size of the ${\cal O}(\alpha)$ electroweak
corrections to the total $p\bar p\to\ell\nu X$ cross section and to 
$R_{W/Z}$ is 
sensitive to the acceptance cuts and whether lepton identification
requirements are taken into account or not. In Table~\ref{TAB:TWO}, we 
list the electroweak $K$-factor,
\begin{equation}
K^{EW}={\sigma^{{\cal O}(\alpha^3)}(p\bar p\to
W\to\ell\nu X)\over\sigma^{\rm Born}(p\bar p\to W\to\ell\nu)}~, 
\end{equation}
and the correction factor for $R_{W/Z}$,
\begin{equation}
K_R^{EW}={R_{W/Z}^{{\cal O}(\alpha^3)}\over R_{W/Z}^{\rm Born}}~,
\end{equation}
for the acceptance cuts listed in Eqs.~(\ref{eq:lepcut})
and~(\ref{eq:ptmisscut}) with and without taking the lepton identification
requirements of Table~\ref{tab:one} into account. As before, we 
include photon exchange and 
$\gamma Z$ interference effects, and impose a cut on the di-lepton 
invariant mass of $75~{\rm GeV}<m(\ell^+\ell^-)<105~{\rm 
GeV}$~\cite{BKS}, in the calculation of the ${\cal O}(\alpha^3)$ $Z$ 
boson cross section entering $R_{W/Z}$. It should be noted that the 
missing purely weak corrections in the calculation of Ref.~\cite{BKS} 
introduce an uncertainty of ${\cal O}(\alpha/\pi)$ in 
$K_R^{EW}$ which could be significant. 

From the results listed in Table~\ref{TAB:TWO} we see that the ${\cal
O}(\alpha)$ electroweak corrections decrease the $W$ cross section and 
increase $R_{W/Z}$ by several per cent for the cuts imposed. As for the
differential cross section, the ${\cal O}(\alpha)$ corrections are 
larger in the electron case when lepton identification requirements are
not included. When lepton identification requirements are included, the
corrections are reduced in the electron case and enhanced in the muon
case. If no acceptance cuts and no lepton identification 
requirements are taken into account, all mass singular terms 
cancel in the total cross section, and the size of the electroweak 
corrections is reduced to about $-0.2\%$. 

The size of the ${\cal O}(\alpha)$ electroweak corrections should be 
compared with that of the ${\cal O}(\alpha_s)$ and ${\cal
O}(\alpha_s^2)$ QCD corrections. NLO QCD corrections are known (see 
{\it e.g.} Ref.~\cite{reno} and references therein)
to enhance the $W$ production cross section by about $15-20\%$ and are
thus significantly larger than the ${\cal O}(\alpha)$ EW
corrections. In fact, the size of the ${\cal O}(\alpha)$ electroweak 
corrections to the $W$ cross section when cuts are imposed 
is about equal to that of the next-to-next-to-leading order (NNLO) QCD 
corrections~\cite{willy}. On the other hand, the ${\cal O}(\alpha)$ 
electroweak corrections to $R_{W/Z}$ are in some cases considerably 
larger than the 
NLO QCD corrections. Since the QCD corrections to $W$ and $Z$ production
are very similar, they cancel almost perfectly in the $W$ to $Z$ cross
section ratio; the ${\cal O}(\alpha_s)$ corrections to $R_{W/Z}$ are of
${\cal O}(1\%)$ or less, depending on the set of parton distribution
functions used~\cite{willy}. In contrast, the electroweak corrections do
in general not cancel in $R_{W/Z}$. As noted before, in
$Z\to\ell^+\ell^-$ both leptons can emit photons, whereas only the
charged lepton radiates in $W\to\ell\nu$ decays. Since final state
photonic corrections are the dominating contribution to the ${\cal
O}(\alpha)$ EW corrections, the ${\cal O}(\alpha)$ corrections to the
$W$ and $Z$ cross sections are quite different, and thus do not 
cancel in $R_{W/Z}$. For example, when lepton identification 
requirements are taken into account, the ${\cal O}(\alpha)$ EW
corrections in the electron (muon) case increase $R_{W/Z}$ by $0.2\%$
($6.5\%$). Note that, unlike the electroweak 
corrections, QCD corrections are 
only slightly modified by cuts and lepton identification requirements. 

From the results shown in Table~\ref{TAB:TWO} we conclude that it 
will be necessary to correct for higher 
order electroweak effects if one wishes to measure the $W$ cross 
section and $R_{W/Z}$ with an accuracy of ${\cal O}(1\%)$ or better.

\subsection{Electroweak Corrections to the Charge Asymmetry of Leptons
in $W$ decays} 

\begin{figure}
\vskip 12cm
\includegraphics{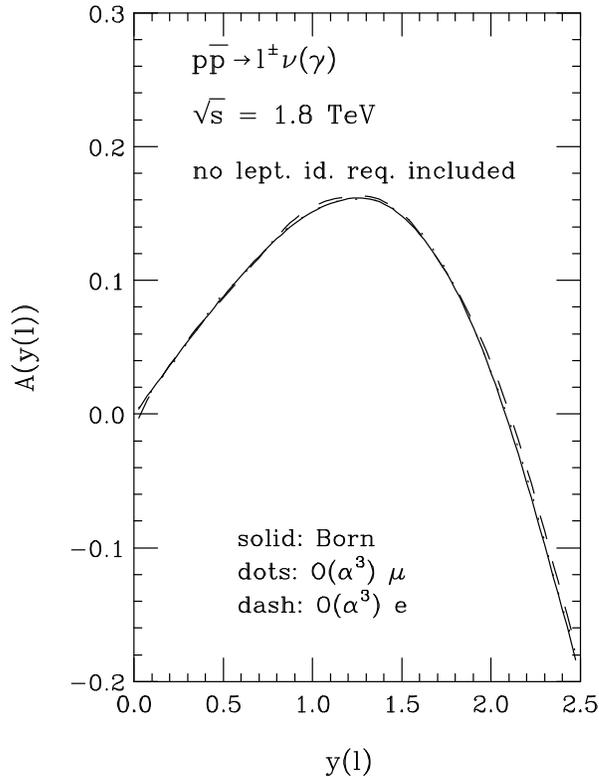}
\vskip -1.cm
\caption{The charge asymmetry for leptons, $A(y(\ell))$, in $W\to\ell\nu$
decays for $p\bar p$ collisions at $\protect{\sqrt{s}=1.8}$~TeV. The dashed 
(dotted) lines show the asymmetry for electron (muon) final states 
including ${\cal O}(\alpha)$ electroweak corrections. The solid line 
gives the Born prediction of $A(y(\ell))$. Except for the 
pseudo-rapidity cut on the charged lepton, the cuts listed in
Eqs.~(\protect{\ref{eq:lepcut}}) and~(\protect{\ref{eq:ptmisscut}}) are
imposed. The lepton identification requirements of
Table~\protect{\ref{tab:one}} are not taken into account. The energy and
momentum resolutions used are described in Sec.~IIC.}
\label{fig:twelve}
\end{figure}

Uncertainties in the parton distribution functions are a major
contribution to the systematic error of the $W$ mass extracted in hadron
collider experiments~\cite{cdfwmass,D0Wmass}. Measurement of the charge
asymmetry of leptons in $W$ decays~\cite{berger},
\begin{equation}
A(y(\ell))={d\sigma^+/dy(\ell)-d\sigma^-/dy(\ell)\over
d\sigma^+/dy(\ell)+d\sigma^-/dy(\ell)}~, 
\end{equation}
where $y(\ell)$ is the lepton rapidity and 
\begin{equation}
\sigma^\pm=\sigma(p\bar p\to\ell^\pm\nu X),
\end{equation}
provides strong constraints on the ratio of $d$ and $u$ quark
distributions~\cite{cdfwasym}. These constraints considerably reduce the
uncertainty originating from the parton distribution functions in the
$W$ mass measurement~\cite{cdfwmass,D0Wmass}. It is thus important to
know how electroweak radiative corrections affect $A(y(\ell))$. The ${\cal
O}(\alpha^3)$ asymmetry as a function of the lepton rapidity for
$e\nu(\gamma)$ (dashed line) and $\mu\nu(\gamma)$ (dotted line)
production is shown in Fig.~\ref{fig:twelve} together with the lowest 
order prediction (solid line). Except for the 
pseudo-rapidity cut on the charged lepton, we impose the cuts listed in 
Eqs.~(\ref{eq:lepcut}) and~(\ref{eq:ptmisscut}) in this subsection. 
Since $A(-y)=-A(y)$, the asymmetry is only displayed for $y(\ell)>0$. 
The flavor specific lepton identification requirements of 
Table~\ref{tab:one} are not taken into account in
Fig.~\ref{fig:twelve}. The 
asymmetry in the Born approximation for electron and muon
final states is virtually indistinguishable for the cuts and the energy 
and momentum resolutions we use. Electroweak corrections are seen to only 
slightly affect the charge asymmetry. 

\begin{figure}
\vskip 12cm
\includegraphics{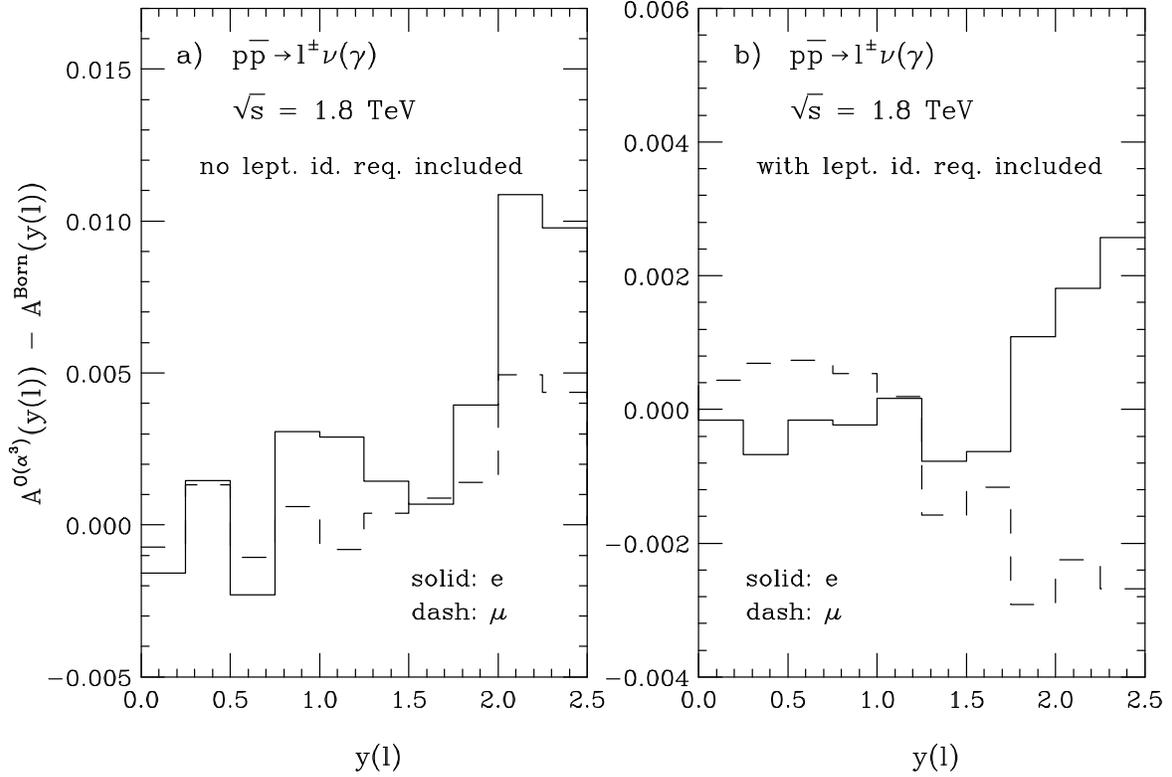}
\vskip -1.cm
\caption{The difference of the ${\cal O}(\alpha^3)$ and the Born charge
asymmetry for electrons (solid) and muons (dashed) a) without and b)
with the lepton identification requirements of Table~\ref{tab:one} 
taken into account. Except for the pseudo-rapidity cut on the charged 
lepton, the cuts listed in Eqs.~(\protect{\ref{eq:lepcut}}) 
and~(\protect{\ref{eq:ptmisscut}}) are imposed. The energy and momentum 
resolutions used are described in Sec.~IIC.}
\label{fig:thirteen}
\end{figure}

In order to display the effect of EW radiative corrections on
$A(y(\ell))$ more clearly, we show the difference between the ${\cal 
O}(\alpha^3)$ and the Born asymmetry in Fig.~\ref{fig:thirteen}.
Without taking the lepton identification requirements of
Table~\ref{tab:one} into account, the difference of the ${\cal
O}(\alpha^3)$ and the Born asymmetry is positive and gradually increases
with $y(\ell)$ from zero at $y(\ell)=0$ to about 0.01 for electrons,
and approximately 0.005 for muons, at $y(\ell)=2.5$ (see 
Fig.~\ref{fig:thirteen}a). Figure~\ref{fig:thirteen}b displays the
difference between the ${\cal O}(\alpha^3)$ and the Born charge
asymmetry when the lepton identification requirements are
included in the simulation. Due to the recombination of the 
electron and photon momentum four vectors for small lepton -- photon 
opening angles, the size of the electroweak corrections to the charge 
asymmetry in the electron case is drastically reduced when these 
requirements are taken into account. For $y(e)\leq 1.5$, the ${\cal 
O}(\alpha)$ corrections reduce $A(y(e))$ by a very small amount.  
In the forward rapidity region, $1.7<y(e)<2.5$, the difference 
of the ${\cal O}(\alpha^3)$ and the Born charge asymmetry is positive
and slowly increases, reaching approximately 0.0025
at $y(e)=2.5$. For comparison, the statistical error in $A(y(e))$
in this region expected for Run~II (assuming $\int\!{\cal L}dt=2~{\rm 
fb}^{-1}$) is $\delta A(y=2.5)\approx 0.005$~\cite{d0upgr,cdfii}. The
variation of the charge asymmetry due to the uncertainties of the
present parton distribution functions is about 0.015~\cite{GK1} in the
same region. The magnitude of the ${\cal O}(\alpha)$ electroweak 
corrections for muons and electrons when lepton identification
requirements are included is similar. However, in the muon
case, the EW corrections enhance (reduce) the ${\cal O}(\alpha^3)$ 
asymmetry at small (large) rapidities.

\section{Conclusions}

The mass of the $W$ boson is one of the fundamental parameters of the
SM and a precise measurement of $M_W$ is an important objective for
current experiments at LEP2 and future experiments at the Tevatron. A
precise measurement of the $W$ mass helps to constrain the Higgs boson 
mass from radiative corrections. It will also provide restrictions on the
parameters of the MSSM. In order to perform such a measurement at a
hadron collider, it
is crucial to fully control higher order QCD and electroweak corrections. 
In this paper we have presented a calculation of the electroweak
corrections to $W$ production in hadronic collisions which is based on 
the full set of contributing ${\cal O}(\alpha^3)$ Feynman diagrams. 

The ${\cal O}(\alpha)$ electroweak corrections can be arranged into
separately gauge invariant QED-like contributions corresponding to
initial state, final state and interference corrections, and gauge
invariant modified weak contributions to the $W$ production and decay
processes. Due to mass singular logarithmic
terms associated with final state photon radiation in the limit where
the photon is collinear with one of the leptons, final state radiation
effects dominate. Initial state corrections were found to be small
after appropriately factorizing the corresponding collinear 
singularities into the parton distribution functions. However, currently
no parton distribution functions which include QED corrections are 
available. With the factorization scheme used in this paper, the effect
of the QED corrections on the PDF is expected to be small. We find that 
the part of the initial state corrections included in our calculation is
uniform over the entire range of the $\ell\nu$
transverse mass and the lepton $p_T$ range, and increases the differential
cross section by about $1\%$. Likewise, the modified weak
corrections are uniform, but decrease the cross section by a similar
amount. In contrast, the final state QED-like corrections modify the 
shape of the $M_T$ and $p_T$ distributions substantially.

Without including the lepton identification requirements imposed by
experiments, the
effect of the electroweak corrections is larger in the electron channel 
than in the muon channel. When these requirements are taken into
account, the mass 
singular logarithmic terms are eliminated in the electron case because 
the electron and photon momentum four vectors are combined for small 
opening angles where it is difficult to resolve the two particles. Initial
state QED-like, final state QED-like, modified weak and interference
contributions are then all of similar size. On the other hand, in order 
to experimentally identify muons, the energy of the photon is required
to be smaller than a critical value if the $\mu-\gamma$ separation is 
small, and mass singular terms survive. Removing energetic
photons thus enhances the effect of the ${\cal O}(\alpha)$ corrections, 
and the effect of the electroweak corrections in the muon case is larger
than in the electron case once lepton identification requirements are
included.

Electroweak radiative corrections have a significant impact on the $W$
mass extracted from experiment. The main effect is caused by final 
state photon radiation and is corrected for in the $W$ mass analyses of
the Tevatron experiments~\cite{cdfwmass,D0Wmass}. However, in the 
calculation used by CDF and D\O~\cite{BK,RGW}, the effect of the
final state soft and virtual photonic corrections is estimated
indirectly from the inclusive ${\cal O}(\alpha^2)$ $W\to\ell\nu(\gamma)$
width and the hard photon bremsstrahlung contribution. Initial state, 
interference, and weak contributions to the ${\cal O}(\alpha)$ 
corrections are ignored altogether. The correct treatment of the final 
state soft and virtual photonic corrections significantly changes the
slope of the transverse mass distribution in the region $M_T>M_W$. This
changes the $W$ mass extracted from the transverse mass
distribution by ${\cal O}(10~{\rm MeV})$, and might also 
have a non-negligible effect on the $W$ width measured from the tail of
the $M_T$ distribution. More detailed numerical simulations are needed to 
quantitatively assess this effect. Initial state, and initial -- final 
state interference corrections, have only a small effect on the $M_T$
distribution and hence are expected to only marginally influence the amount
the $W$ boson mass is shifted. 

Our results demonstrate that, for the current level of precision, the 
approximate 
calculation of Ref.~\cite{BK} is adequate. The small difference in the
$W$ boson mass obtained in the complete ${\cal O}(\alpha^3)$ and the
approximate calculation, however, 
cannot be ignored if one attempts to measure the $W$ mass with high
precision at hadron colliders. This also raises the question of how
strongly multiple final state photon radiation influences the measured $W$
boson mass. So far, only partial calculations 
exist~\cite{Was}. A more complete understanding of multiple photon
radiation is warranted. 

As an alternative to the transverse mass and the lepton $p_T$
distribution, the $W$ to $Z$ transverse mass ratio, $R_{M_T}$, has been 
used recently to extract the mass of the $W$ boson. We found that, since
the ${\cal O}(\alpha)$ corrections to the $Z$ boson transverse mass 
distribution are significantly larger than those to the $W$ $M_T$ 
distribution, electroweak corrections influence $R_{M_T}$ more strongly 
than the $M_T$ or $p_T(\ell)$ distribution. 

Finally, we studied how electroweak radiative corrections influence the
$W$ cross section, the $W$ to $Z$ cross section ratio, $R_{W/Z}$, and 
the charge asymmetry of leptons in $W$ decays, $A(y(\ell))$. 
As shown in Table~\ref{TAB:TWO}, the ${\cal O}(\alpha)$ electroweak 
corrections can reduce the $W$ cross section by up to $5\%$ in the
presence of cuts. The size of the ${\cal O}(\alpha)$ electroweak 
corrections is of the same order as the NNLO QCD
corrections~\cite{willy}. The ${\cal O}(\alpha)$
corrections were found to enhance $R_{W/Z}$ by up to $6.5\%$. QCD 
corrections, on the
other hand, cancel almost perfectly in the $W$ to $Z$ cross section
ratio and are of ${\cal O}(1\%)$. The ${\cal O}(\alpha)$ electroweak 
corrections to $A(y(\ell))$ may not be negligible in view of the 
projected accuracy of the charge asymmetry in future Tevatron runs.

\acknowledgements
We would like to thank I.~Adam, M.~Demarteau, S.~Errede, E.~Flattum,
H.~Frisch, U.~Heintz, Y-K.~Kim and E.~Laenen for stimulating discussions.
Two of us (U.B. and D.W.) are grateful to the Fermilab Theory Group,
where part of this work was carried out, for its generous hospitality.
This work has been supported in part by Department of Energy 
contract No.~DE-AC02-76CHO3000 and NSF grant PHY-9600770. 

%
%

%
\newpage
%
\widetext
\begin{table}
\caption{Summary of lepton identification requirements.} 
\label{tab:one}
\vskip 5.mm
\begin{tabular}{cc}
\multicolumn{1}{c}{electrons} & \multicolumn{1}{c}{muons} \\
\tableline
combine $e$ and $\gamma$ momentum four vectors if & reject events with 
$E_\gamma>2$~GeV \\ 
$\Delta R_{e\gamma}<0.2$ and if $E_\gamma<0.15~E_e$ for 
$0.2<\Delta R_{e\gamma}<0.3$ & for $\Delta R_{\mu\gamma}<0.2$ \\
\tableline
reject events with 
$E_\gamma>0.15~E_e$ & reject events with 
$E_\gamma>6$~GeV \\  
for $0.2<\Delta R_{e\gamma}<0.4$ & for $0.2<\Delta R_{\mu\gamma}<0.6$ \\
\end{tabular}
\end{table}
\vskip 5.mm
\begin{table}
\caption{The electroweak $K$-factor $K^{EW}=\sigma^{{\cal
O}(\alpha^3)}(p\bar p\to W\to\ell\nu X)/\sigma^{\rm Born}(p\bar p\to 
W\to\ell\nu)$ ($\ell=e,\,\mu$) and the correction factor to $R_{W/Z}$, 
$K_R^{EW}=R_{W/Z}^{{\cal O}(\alpha^3)}/R_{W/Z}^{\rm Born}$, with 
\protect{$75~{\rm GeV}<m(\ell^+\ell^-)\break <105$}~GeV, for $p\bar p$
collisions at $\protect{\sqrt{s}=1.8}$~TeV. Shown are the predictions 
without and with the lepton identification requirements of
Table~\protect{\ref{tab:one}} taken into account. The cuts imposed are 
listed in Eqs.~(\protect{\ref{eq:lepcut}})
and~(\protect{\ref{eq:ptmisscut}}). The energy and momentum resolutions
used are described in Sec.~IIC. }
\label{TAB:TWO}
\vskip 5.mm
\begin{tabular}{ccc}
\multicolumn{1}{c}{} &
\multicolumn{1}{c}{without lepton id.} &
\multicolumn{1}{c}{with lepton id.} \\
\multicolumn{1}{c}{} &
\multicolumn{1}{c}{requirements} &
\multicolumn{1}{c}{requirements} \\
\tableline
$K^{EW}~(p\bar p\to e^+\nu X)$ & 0.955 & 0.984 \\
$K^{EW}~(p\bar p\to\mu^+\nu X)$ & 0.975 & 0.947 \\
\tableline
$K^{EW}_R~(e)$ & 1.032 & 1.002 \\
$K^{EW}_R~(\mu)$ & 1.012 & 1.065 \\
\end{tabular}
\end{table}

\newpage
%
%
%
\newpage
%
%
%
%
%
\newpage
%
%
\newpage
\newpage

\newpage

\newpage

\newpage

\newpage

\newpage

\newpage

\newpage

\newpage

\end{document}